\documentstyle[aps,prc,twocolumn,floats,epsfig]{revtex}
\draft

\newcommand{\bra}[1]{\langle #1|}
\newcommand{\ket}[1]{|#1\rangle}

\newcommand {\Jcap} {\mathcal J}

\newcommand{\C}{{\sf C\hspace*{-0.9ex}\rule{0.15ex}%
       {1.3ex}\hspace*{0.9ex}}}

\newcommand{\rb}[1]{\raisebox{1.5ex}[0pt]{#1}}

\begin{document}

\renewcommand{\thefootnote}{\arabic{footnote}}
\twocolumn[\columnwidth\textwidth\csname@twocolumnfalse\endcsname
\title{Gamow Shell Model Description of Weakly Bound Nuclei
and Unbound Nuclear States}

\author{N. Michel,$^1$  W. Nazarewicz,$^{2-4}$ M. P{\l}oszajczak,$^1$
and  J. Oko{\l}owicz$^5$}

\address{$^1$GANIL, CEA/DSM-CNRS/IN2P3, BP 5027, F-14076 Caen Cedex 05, France}

\address{$^2$Department of Physics and Astronomy,
              The University of Tennessee,
              Knoxville, Tennessee 37996}
              
\address{$^3$Physics Division,
              Oak Ridge National Laboratory,
              P. O. Box 2008, Oak Ridge, Tennessee 37831}

\address{$^4$Institute of Theoretical Physics,
              Warsaw University,
              ul. Ho\.za 69, PL-00681, Warsaw, Poland}

\address{$^5$Institute of Nuclear Physics, PL-31342 Krak\'ow, Poland}
\date{\today}

\maketitle
\addvspace{5mm}
%
%
\begin{abstract}
We present the   study of weakly bound,  neutron-rich nuclei 
using the nuclear shell model employing  the complex  Berggren ensemble
representing the bound single-particle states, unbound Gamow states,
and the non-resonant continuum. In the proposed Gamow Shell Model, the
Hamiltonian consists of a one-body finite depth (Woods-Saxon) potential and a
residual two-body interaction. We discuss 
the basic ingredients of the Gamow Shell Model. The  formalism
is illustrated by  calculations  involving  {\it
several} valence neutrons outside the double-magic core: $^{6-10}$He and
$^{18-22}$O. 
\end{abstract}
\addvspace{5mm}]
\narrowtext
%
%
\section{Introduction}

The major theoretical challenge in the microscopic
description of weakly bound nuclei
is the rigorous treatment of both the many-body correlations 
and the continuum of positive-energy states and decay channels. 
A fully symmetric
description of the interplay between scattering states, resonances, and bound states in the
many-body wave function requires a close interplay between methods of nuclear
structure and nuclear reactions. This mutual cross-fertilization, 
which cannot be accomplished without overcoming a traditional
separation between  nuclear structure and nuclear reaction methods, is a splendid 
opportunity for  opening a new era
in the nuclear theory of loosely bound systems.

In many respects, weakly bound nuclei
are much more difficult to treat theoretically than well-bound systems \cite{[Dob97a]}.
The major theoretical difficulty and 
challenge is the  treatment of the particle continuum.
For weakly bound nuclei (or for nuclear states above the particle threshold),
the continuum of positive-energy states and resulting decay channels
must be taken into account explicitly.
As a result,
many cherished approaches of nuclear theory such as the
conventional  shell model (based on a single-particle basis of bound states)
 and  the
pairing theory 
must be modified. 

There are many factors which make the coupling to the particle continuum important.
Firstly, even for a bound nucleus, there appears a virtual 
scattering  into the phase space of unbound states. Although this process involves
intermediate scattering states,  the correlated bound states must be particle
stable, i.e., they must
have zero width.
Secondly, the properties of unbound states, i.e., above the particle (or cluster) threshold,
 directly reflect the continuum structure. In addition, continuum coupling directly
  affects   the effective nucleon-nucleon interaction.

The impact  of the particle continuum was discussed in the early days of the 
multiconfigurational shell model (SM)  in the middle of the last century. However,
thanks to  the tremendous success of
the  large-scale SM in terms of  interacting nucleons assumed to be
 {\it perfectly
isolated} from an {\it external environment} of scattering states 
\cite{[Cau02],[Cau02a],[Ots02],[Cor02],[Bro02]}, 
the continuum-related matters had been swept under the rug.
An example of an impact  of the continuum that goes beyond the
standard SM physics
is the so-called Thomas-Ehrman shift \cite{[Tho51],[Ehr51]}
appearing in, e.g., the
mirror nuclei $^{13}$C, $^{13}$N, which is a salient effect of a
coupling to the continuum  depending on the position of the respective particle
emission thresholds. The mathematical formulation of the problem of
nuclear states
embedded in the continuum of decay channels goes back to
Feshbach \cite{[Fes62]}, who
introduced the two subspaces containing the discrete  and 
scattering states.
 Feshbach succeeded in formulating a unified description of nuclear
reactions for  both direct processes in the short-time scale and compound
nucleus processes in the long-time scale. As far as nuclear structure is concerned,
the treatment 
of excited states near or above the decay threshold
has been  a playground of
the continuum shell model (CSM) \cite{[Fan61],[Mah69],[Bar77],[Phi77],[Hal80],[Rot91]}.
Unfortunately, a unified description of nuclear
structure and nuclear reaction aspects is much more complicated and became
possible in realistic situations only at the end of the last century 
(see Ref.~\cite{[Oko03]} for a recent review). 

In the CSM, including the recently developed
 Shell Model
Embedded in the Continuum (SMEC) \cite{[Ben99],[Ben00b],[Ben00c]},
the scattering 
states  and bound states are treated  on an  equal footing.
So far, most applications of the CSM,  including SMEC, 
have been used to describe limiting situations in which there is coupling to
 {\em one-nucleon decay channels} only. However, by allowing  only one particle
to be present in  the  continuum,
it  is impossible to apply 
the CSM to     `Borromean systems'
for which $A$- and $(A$-2)-nucleon systems are particle-stable but the intermediate
$(A$-1)-system is  not.
Various approaches, including  the
hyperspherical harmonic method or the coupled-channel approach,
have been developed to study structure and reaction
aspects of three-body weakly bound nuclei \cite{[Dan98],[Nie01],[Esb99]}. However,  most of these
models utilize  the particle-core  coupling  which does not allow for the exact treatment
of  core excitations  and the antisymmetrization between the core nucleons and the valence
particles.

The reason for limiting oneself to only one particle in the continuum 
in the CSM has been two-fold. First, 
 the number of scattering states
needed to properly describe the underlying dynamics can easily go beyond the limit of
what present computers can handle. Second,  treating the
continuum-continuum coupling, which is always present when
two or more particles are  scattered to unbound levels, is difficult.
There have been 
only a few  attempts to treat the multi-particle case \cite{[Epp75],[Wen87]}
 and, unfortunately,
the proposed numerical schemes, due to their complexity, have never
been adopted in microscopic  calculations involving multiconfiguration mixing.
Consequently,   an entirely different approach is called for.

Recently,  we formulated and tested the multiconfigurational  shell model 
in the complete Berggren basis \cite{[Mic02]}, the so-called Gamow Shell Model (GSM).
 (For application to two-particle
resonant states, see  also Ref. \cite{[Bet02],[Bet02a]}.)
In this paper,  GSM is
applied to  systems containing several  valence neutrons.
The single-particle (s.p.) basis of GSM is given by the Berggren ensemble, which
contains Gamow states and the non-resonant continuum. 
The Gamow states \cite{[Gam28]}
(sometimes called Siegert \cite{[Sie39]} or  resonant states) 
were introduced for the first time in 1928 to study
the $\alpha-$resonances. Gamow defined complex-energy eigenstates
$E = E_0 - i \Gamma/2 $ in order to describe the particle emission in
the quasi-stationary formalism. Indeed, if one looks at the temporal part of such a
state, which is  $e^{iE_0t/ \hbar} e^{-\Gamma t/(2\hbar)}$, one notices 
that the squared modulus of the wave function has the time-dependence 
$\propto e^{-\Gamma t}$, and one can identify $\hbar/(\Gamma \log 2)$ 
with the half-life of the system. 

Formally, the resonant  states
 are  generalized eigenstates  of the time-independent 
Schr\"odinger equation with  purely outgoing boundary conditions.
 They  correspond to
the poles of the $S$-matrix in the complex energy plane  lying on or 
below the positive real axis; they are regular in origin 
  and satisfy purely outgoing  asymptotics. 
In the quasi-stationary approach with Gamow states, each observable $O$ is complex. An 
interpretation of these complex
values has been given by Berggren \cite{[Ber96a]}: the real part of the matrix
element gives the  average value, while 
the imaginary part represents  the uncertainty of the mean value. 
This is due to the finite lifetime of the Gamow state which implies that
none of the measurements in this state can have a well-defined probability.

In the previous pilot work \cite{[Mic02]} we showed first applications of the GSM.
In this work, we give the details of calculations and demonstrate first
applications of GSM to particle distributions and transition matrix elements.

The paper is organized as follows.  Section~\ref{prol_analy} discusses
how to calculate  the matrix elements in the Berggren basis.
The completeness relations valid for  the  single-particle
resonant states are briefly reviewed in Sec.~\ref{complet}, and some
numerical examples involving the Berggren set  of the Woods-Saxon
potential are presented. 
Section~\ref{model} describes the GSM Hamiltonian
used in our work.
The extension of the completeness relations
to the many-body case is
described in Sec.~\ref{completmb}. Sections \ref{chaine_O}
and \ref{chaine_He} contain the GCM analysis of $^{18-22}$O
and $^{6-10}$He, respectively.
Finally, Sec.~\ref{summary} contains the main conclusions of the paper.

\section {Matrix elements in the Berggren basis}
\label{prol_analy}

Gamow functions are solutions of the Schr\"odinger equation
which are regular at the origin and have
the outgoing wave asymptotics; i.e.,  the  radial part behaves as $e^{ikr}$ at large distances.
In the case of the spherical one-body potential, the resonant wave function 
$\phi_{nj\ell}$  carrying the s.p. angular momentum $\ell j$ 
can be written as a product of
the usual angular part and the radial wave function  $u_{nj\ell}(r)/r$.
It is customary to introduce the notation
\begin{eqnarray}
\tilde{u}_{nj\ell}(r) &=&{u}_{nj\ell}(r)^*\\
\tilde{\phi}_{nj\ell} & =& {\phi}_{nj\ell}(u\rightarrow \tilde{u}).  
\end{eqnarray}
For bound states, one can always introduce a
phase convention which makes the radial wave function real.
That is, for bound states $\tilde{\phi}_{nj\ell}={\phi}_{nj\ell}$. 
The following discussion concerns the specific properties of the 
 radial Gamow wave functions $u_{nj\ell}(r)$. Of course, one should always remember
 that the angular part is always present, but its treatment is standard. Consequently,
in the GSM, 
 only  radial matrix elements require special attention.

\subsection {Normalization of  Gamow states and one-body matrix elements}

The norm $N_i$ of a resonant state,
\begin{equation}\label{Gamnorm}
N^2_i=\int_{0}^{+\infty} \!\!\!\! u^2_i(r) dr,
\end{equation}
 and the radial matrix 
elements calculated in the Berggren basis,
\begin{equation}\label{Gammat}
O_{if}=\int_{0}^{+\infty} \!\!\!\! u_f(r) \; O(r) \; u_i(r) \, dr, 
\end{equation}
  are diverging, but this
difficulty can be avoided by means of a regularization procedure 
\cite{[Zel60],[Hok65],[Rom68],[Ber68],[Gya71],[Gar76]}.
Zel'dovich proposed to
multiply the integrand of a radial matrix element by a Gaussian 
convergence  factor  \cite{[Zel60]}:
\begin{eqnarray}
\label{z1}
\langle u_f|O|u_i\rangle = \lim_{\epsilon \to 0} \int_{0}^{+\infty} \!\!\!\!
e^{-\epsilon r^2} u_f(r) \; O(r) \; u_i(r) \, dr.
\end{eqnarray}
In this expression, $|u_f\rangle$ and $|u_i\rangle$ stand
for  single-particle (s.p.)  states, and
$O(r)$ is a radial part of a one-body operator. Using the Zel'dovich regularization method, 
Berggren has shown \cite{[Ber68]}   that the Gamow states,
together with scattering states,   form a complete basis. In particular, 
using  definition 
(\ref{z1}) with $O(r)$=1, one can demonstrate that all Gamow states can be orthonormalized, and
such orthonormalized functions can  
be used to calculate  matrix elements.
Unfortunately, the method of Zel'dovich, even though important on  formal
grounds, cannot be used in numerical applications due to the difficulty in
approaching the limit in (\ref{z1}) for diverging integrals.

\begin{figure}[htb]
  \begin{center}
 \leavevmode
  \epsfxsize=7.5cm
  \epsfbox{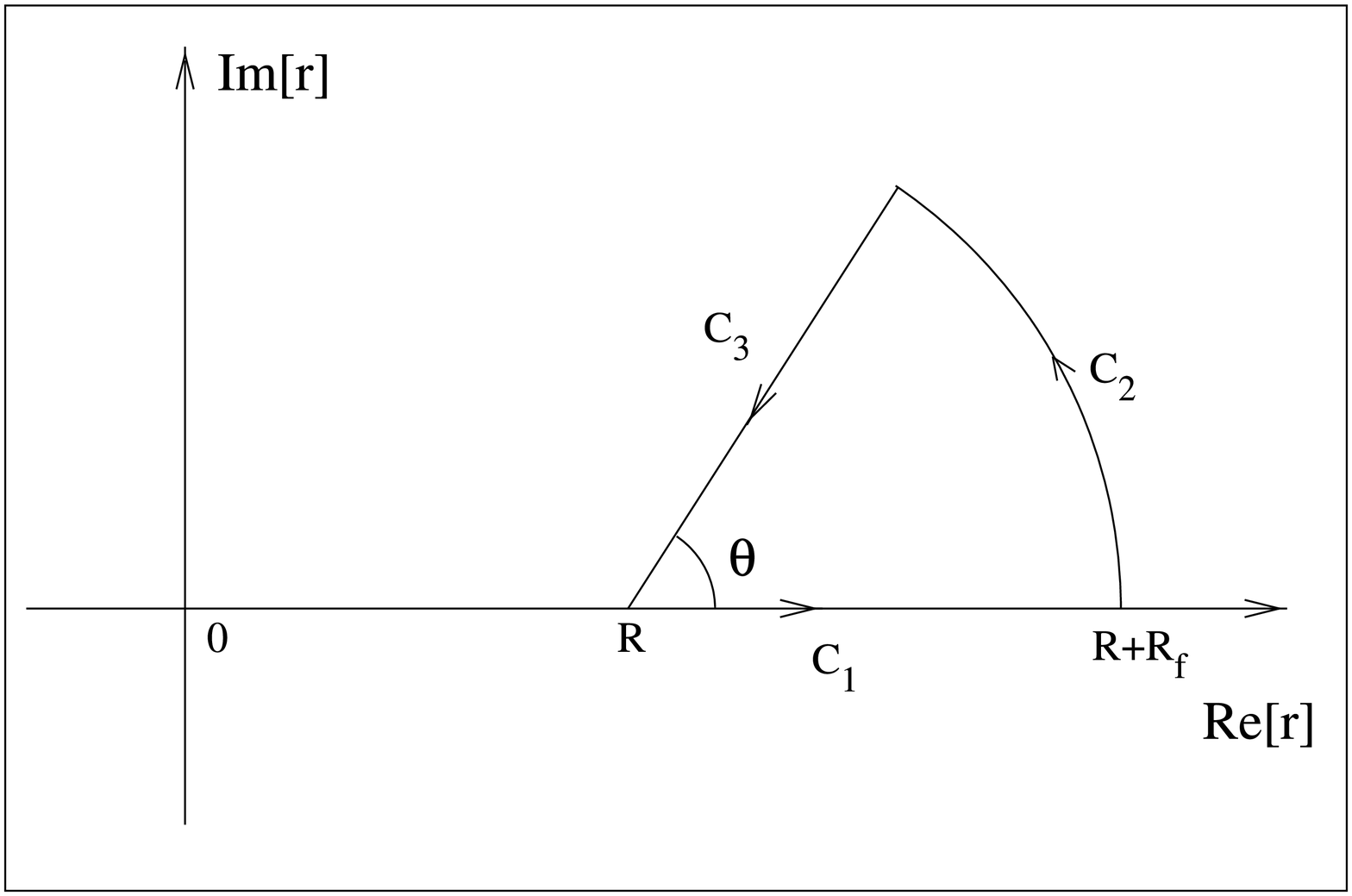}
\caption{The path in the complex coordinate space corresponding to the 
complex rotation by angle $\theta$. $R$ is the point from which the exterior
complex rotation starts. $R$ is large as compared to the nuclear radius;
hence it is assumed that  the nuclear potential is negligible for $r$$>$$R$.
}
\label{ext_cs}
\end{center}
\end{figure}

An equivalent and more practical procedure, justified by  the apparatus of the 
analytical continuation, was proposed by Gyarmati and Vertse \cite{[Gya71]}. For that,
let us define the following functional on $(-\infty ; V_{lim})$ \cite{[Kuk89]}:
\begin{eqnarray}
\label{z2}
F(V_o) = \frac{O_{ij}}{N_iN_f},
\end{eqnarray}
where $V_o$ is the depth of the potential generating s.p.
wave functions $u_f$ and $u_i$, $V_{lim}$ is the depth of the potential for
which one of these functions is bound and the other one is at zero energy,
and $O(r)$ is some analytical operator. This functional is defined 
in such a way because the
integral converges in the domain $(-\infty ; V_{lim})$. It represents the
radial matrix element $\langle u_f|O|u_i\rangle/||u_f||/||u_i||$ between two
not necessarily normalized wave functions. Since   $u_f$, $u_i$ are bound,
one can make them real, $u^{*}_f = u_f$, $u^{*}_i = u_i$.
 In Ref.~\cite{[Kuk89]}, the analytical continuation of $F$ is made using
the Pad{\'e} approximants. In the present work we shall 
refer to  the technique of the
complex rotation \cite{[Gya71]} which allows calculation of $F$ with $V_o > V_{lim}$. To
see that, let us call $f(r)$ one of three integrands $u_f(r) O(r) u_i(r)$, 
$u^2_f(r)$, or $u^2_i(r)$ and let us take $V_o < V_{lim}$. Since $f$
is analytical on $\C$ (see Fig. \ref{ext_cs}), then, 
following the Cauchy theorem, one has:
\begin{eqnarray}
\int_{C_1} f(z) \, dz &+& \int_{C_2} f(z) \, dz \nonumber
\\
 & + &  \int_{C_3} f(z) \, dz = 0.
\end{eqnarray}

Since $f$ decreases exponentially for Re$[z] > 0$, the integral $\int_{C_2}
f(z) \, dz \to 0$ if $R_f \to +\infty$. For the same reason, the integrals 
$\int_{C_1} f(z) \, dz$ and $\int_{C_3} f(z) \, dz$ converge if $R_f \to
+\infty$. Consequently, for $R_f \to +\infty$ one obtains:
\begin{eqnarray}
\label{z3}
\int_{R}^{+\infty} f(r) \, dr = \int_{0}^{+\infty} \!\!\!\! f(R +
x\cdot e^{i\theta}) e^{i\theta} \, dx.  
\end{eqnarray}

Hence, on the interval $(-\infty ; V_{lim})$,  one can define $F$ by 
Eq.~(\ref{z2}), with the norm (\ref{Gamnorm}) given by
\begin{eqnarray}
\label{z6}
N_i &=& \sqrt{\int_{0}^{R} u^2_i(r) \, dr +  \int_{0}^{+\infty} \!\!\!\!
u^2_i(R + x\cdot e^{i\theta}) \; e^{i\theta} \, dx} 
\end{eqnarray}
and with the matrix element 
(\ref{Gammat}) of the form:
\begin{eqnarray}
\label{z5}
O_{if} & = & \int_{0}^{R} u_f(r) \; O(r) \; u_i(r) \, dr\nonumber  \\
  & + & \int_{0}^{+\infty} \!\!\!\!\left[ u_f(R + x\cdot e^{i\theta}) \; O(R + x\cdot
e^{i\theta}) \right. \\
 & \strut &
\strut~~~~~~~~~\left.\times u_i(R + x\cdot e^{i\theta}) e^{i\theta} \, dx\right].  \nonumber
\end{eqnarray}

If $u_f$ and $u_i$ are bound- or decaying-state  wave functions, one can write 
$k_f = |k_f| e^{-i \alpha_f }$ and $k_i = |k_i| e^{-i \alpha_i }$. 
As $u_f(z) \sim  a_f(z)
e^{ik_fz}$ and $u_i \sim a_i(z) e^{ik_fz}$ when Re$[z] \to +\infty$,  with $a_f$
and $a_i$ the algebraic increasing functions, the integrals defining 
$F$ converge if one takes
\begin{equation}
 $$\theta > \alpha_f + \alpha_i.
  \label{condo_conv}
\end{equation}
In addition, the expression for $F$ is analytical
 because  $F$ is a function of converging
integrals of analytical functions. Square
roots in Eq.~(\ref{z6}) cause no problems because $N^2_i$ and $N^2_f$ have always
a positive real part. 
Consequently, following the theorem of
analytic continuation, Eq.~(\ref{z2}) defines also  $F$ for $V > V_{lim}$. 
In this way, one may calculate  
the radial matrix elements of resonance states which {\it a priori} are not
normalizable.

\subsection {Scattering states}

Scattering  states represent the non-resonant continuum
and explicitly enter the completeness relations discussed
in Sec~\ref{complet}.  Their asymptotic
behavior at $r \to +\infty$ is:
\begin{equation}
\label{s1}
u(r) \sim C^{+} H^{(+)}_{\ell,\eta} (kr) + C^{-} H^{(-)}_{\ell,\eta} (kr),
\label{condition_en_infini_scat} 
\end{equation}
where $H^{(\pm)}_{\ell,\eta}$ denote 
Hankel (or Coulomb) functions. As usual, $u(r)$ 
is normalized to the Dirac $\delta-$distribution:
\begin{equation}
\label{norme_delta}
\int_{0}^{+\infty} \!\!\!\! \tilde{u}(k,r) \; u(k',r) \, dr = \delta(k - k'),
\end{equation}
which gives
\begin{equation}
\label{s3}
C^{+} C^{-} = \frac{1}{2 \pi}.
\end{equation}
Knowing $u(r)$, $H^{(+)}_{\ell,\eta}$, $H^{(-)}_{\ell,\eta}$, and their derivatives
at  point $R$, one may determine  coefficients $C^{+}$ and $C^{-}$ up to
a normalization factor by solving the set of linear equations:
\begin{eqnarray}
\label{s4}
u(R) &=& C^{+}  H^{(+)}_{\ell,\eta} (kR) + C^{-}  H^{(-)}_{\ell,\eta} (kR)  \nonumber
\\     \\
u'(R)&=& k C^{+} \left[ \frac{dH^{(+)}_{\ell,\eta}}{dz} \right]_{z = kR} 
       + k C^{-} \left[ \frac{dH^{(-)}_{\ell,\eta}}{dz} \right]_{z = kR}. \nonumber
\end{eqnarray}
Finally,  the scattering state is normalized to satisfy the condition (\ref{s3}).

\subsection {Matrix elements involving scattering states}

The calculation of matrix elements involving the scattering states is based on the
complex rotation  (\ref{z6},\ref{z5}). However, the analytical continuation
should be introduced differently than  for resonant states. The one-body
matrix element can be written as:
\begin{eqnarray}
\label{s5}
F(k_f) & = & \int_0^{R} \!\!\!\! u_f(r) V(r) u_i(r) \, dr + 
A_f A_i F_{++}(k_f) \nonumber \\
&+& A_f B_i F_{+-}(k_f) 
+ B_f A_i F_{-+}(k_f) \\
&+& B_f B_i F_{--}(k_f), \nonumber
\end{eqnarray}
where
\begin{itemize}
\item $u_f = A_f u^+_f + B_f u^-_f$;
\item $u_i = A_i u^+_i + B_i u^-_i$, where $k_i$ in $F$ is fixed and, 
in general, $u_i$ can be either a bound, resonant or scattering state;
\item $\displaystyle F_{s_f s_i}(k_f) = \int_{0}^{+\infty} u^{s_f}_f(R+x) O(R+x) u^{s_i}_i(R+x) dx $
      with $s_f, s_i \in (+,-)$.	
\end{itemize}
This separation is necessary because the presence of incoming and outgoing
waves in the same integral does not allow one to find a unique path in the complex plane
along which the integrand decreases exponentially. Consequently, for each $F_{s_f s_i}$  
one has to consider the domain of the complex plane where it converges, and 
then one performs an analytical continuation with 
the appropriate angle $\theta_{s_f s_i}$.

Certain integrals cannot be regularized in the above sense. Those include 
$F_{+-}$ and $F_{-+}$ with  $u_i = u_f$. For $O(r) = 1$, 
the integrand tends toward a constant value at $+\infty$, independently of the value
$\theta_{+-}$. This can be immediately seen for neutrons, because with
$z=R+x\cdot e^{i\theta}$ and with  $|z|\rightarrow +\infty$, the product
$u^+(z) \cdot u^-(z) \to \mbox{const} \times e^{ikz} \times e^{-ikz} = 
\mbox{const}$, and the corresponding integral diverges. In this case, 
however, it is easy to see
that the integral is in fact a $\delta-$
distribution, and  it can be calculated by using
 a discrete representation of the Dirac $\delta-$function,
\begin{eqnarray}
\delta(k-k_0) \to \frac{\delta_{k,k_0}}{\Delta k} \label{delta_discret},
\end{eqnarray}
with  $\Delta k$ being the discretization step in $k$.

\section {Completeness relation involving  single-particle Gamow states}
\label{complet}

There exist several  completeness relations involving resonant states. As 
shown by   Lind \cite{[Lin93]}, they all can be derived from Mittag-Leffler theory.
In the following, we briefly discuss the Berggren completeness relation
\cite{[Ber68]} which is used in our paper.
The following discussion will concern the s.p. radial wave functions
corresponding to a given partial wave $(j,\ell)$.  

We begin from the completeness relation of Newton
\cite{[New82]}:
\begin{eqnarray}
\label{x1}
\sum_{n} |u_n\rangle \langle u_n| +  \int_{0}^{+\infty} |u_k\rangle 
\langle u_k| \, dk = 1, 
\end{eqnarray}
where $|u_n\rangle$ are the normalized bound states and $|u_k\rangle$ are the
scattering states along the real energy axis normalized according to (\ref{norme_delta}). 
In the  basis (\ref{x1}), one can expand any bound state or scattering state with 
{\it real energy}. Unfortunately, in the presence of narrow resonances, the 
discretization of the real energy continuum becomes cumbersome. The complex-energy
formalism of Gamow states offers
a simple remedy to this difficulty.

\begin{figure}[htb]
  \begin{center}
 \leavevmode
  \epsfxsize=7.5cm
  \epsfbox{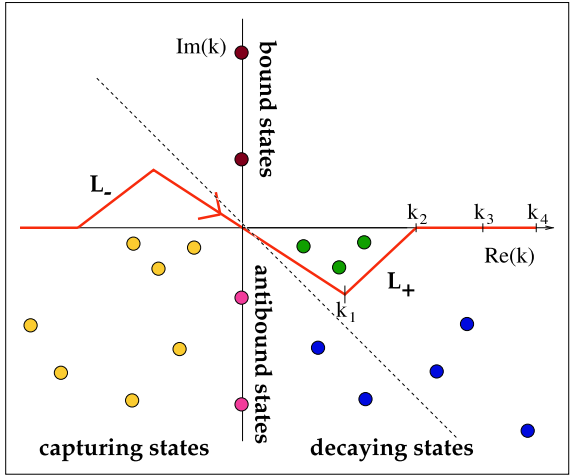}
\caption{Representation of the complex $k-$plane, showing the positions of
bound states, resonances, antiresonances, and the antibound states. $L_{+}$ is
the contour representing
 the non-resonant continuum. The Berggren completeness relation involves  
 the bound states, decaying states lying between $L_{+}$ and real $k-$axis, and
the scattering states on $L_{+}$.
The contour $L_+$  has to be chosen in such a way that all the
poles 
in the discrete sum in Eq.~(\protect\ref{Berggren_comp}) are contained in the 
domain between $L_+$ and the
real energy axis. 
}
\label{Gamow_basis}
\end{center}
\end{figure}

In order to derive the completeness relation with Gamow states, 
one has to deform the
integration contour into the complex $k-$plane, as shown
in  Fig. \ref{Gamow_basis}. 
Following the residuum theorem, one obtains:
\begin{eqnarray}
\label{x2}
-\int_{0}^{+\infty} |u_k\rangle \langle u_k| \, dk &+ & \int_{L_+} |u_k\rangle 
\langle\widetilde{u_k}| \, dk \nonumber \\
&=&  2i\pi \sum_{k_n} Res \left( |u_k\rangle \langle\widetilde{u_k}| 
\right)_{k=k_n},
\end{eqnarray}
where $k_n$ are the poles of $|u_k\rangle \langle\widetilde{u_k}|$ 
lying between the
real axis and the complex contour. 

In general, the scattering wave function $u_k$ can be written as :
\begin{eqnarray}
\label{x3}
u_k(r) = \sqrt{\frac{-{\Jcap}^-(k)}{2\pi {\Jcap}^+(k)}} u^+_k(r) + 
\sqrt{\frac{-{\Jcap}^+(k)}{2\pi {\Jcap}^-(k)}} u^-_k(r).
\end{eqnarray}
In the above expression, ${\Jcap}^{\pm}$ stands for the Jost function \cite{[Jos47]}:
\begin{eqnarray}
\label{J+} 
{\Jcap}^{\pm}(k) = u^{\pm} \frac{du}{dr} - u \frac{du}{dr}^{\pm},
\end{eqnarray} 
where $u^{\pm} \sim H^{\pm}_{l,\eta}(kr)$ when $r \to +\infty$, and  
$u \equiv Au^+ + Bu^-$. For bound and decaying states,
${\Jcap}^-(k)$$\neq$0 (${\Jcap}^-(k) = 0$ only for capturing
states with the 
incoming wave asymptotics). Consequently, the pole of $u_k(r)$
corresponds to zero of ${\Jcap}^+(k)$.
When $k \to k_n$, {\it i.e.}, when $u_k$ 
approaches the resonant state, then:
\begin{eqnarray}
\label{x4}
|u_k\rangle \langle\widetilde{u_k}| \sim -\frac {{\Jcap}^-(k_n)}
{2\pi {\Jcap}^+(k)} |u^+_{k_n}\rangle \langle\widetilde{u^+_{k_n}}|.
\end{eqnarray}
The derivative of the Jost function at $k=k_n$ is
\begin{eqnarray}
\label{x5}
\left[ \frac{d{\Jcap}}{dk} \right]_{k = k_n} = i {\Jcap}^-(k_n) 
Reg \left[ \int_{0}^{+\infty} \!\!\!\! {u^+_{k_n}}^2(r) \, dr \right].
\end{eqnarray}
As this derivative is nonzero, we have as $k \to k_n$:
\begin{eqnarray}
\label{x6}
{\Jcap}^+(k) \sim (k - k_n) \left[ \frac{d{\Jcap}^+}{dk} \right]_{k = k_n},
\end{eqnarray}
hence
\begin{eqnarray}
\label{x7}
|u_k\rangle \langle\widetilde{u_k}| \sim -\frac {1}{2\pi i(k-k_n)} |u_{n}\rangle 
\langle\widetilde{u_{n}}|,
\end{eqnarray}
where  the normalized resonant state is:
\begin{eqnarray}
u_n(r) = u^+_{k_n}(r) \left( Reg \left[ \int_{0}^{+\infty} \!\!\!\! 
{u^+_{k_n}}^2(r) \, dr \right] \right)^{-\frac{1}{2}}.
\label{Gamow_norm}	
\end{eqnarray}
 Finally,  the residuum at  $k=k_n$ is
\begin{eqnarray}
\label{x8}
Res \left( |u_k\rangle \langle\widetilde{u_k}| \right)_{k=k_n} = 
-\frac {1}{2i\pi}  |u_n\rangle \langle\widetilde{u_n}|		     	
\end{eqnarray}
and the completeness relation follows immediately:
\begin{eqnarray}
\sum_{n} |u_n\rangle \langle\widetilde{u_n}| +  
\int_{L_+} |u_k\rangle \langle\widetilde{u_k}| \, dk = 1.
\label{Berggren_comp}  
\end{eqnarray}
In the above equation  $|u_n\rangle$ are the Gamow states (both
 bound states and the decaying resonant states
between the real $k-$axis and the complex contour). 
Relation (\ref{Berggren_comp})
is the Berggren completeness relation which allows one to expand the 
states with complex $k$ inside
the zone between real $k-$axis and the complex contour. 
One may notice again that the resonances in Eq.~(\ref{Gamow_norm}) are normalized
using the squared wave function  and not the modulus of the squared wave function.
This is a consequence of the analytical continuation 
which is used to introduce the normalization of Gamow states.

Figure \ref{Gamow_basis} illustrates the ingredients entering
Eq.~(\ref{Berggren_comp}). The resonant states, the poles of the $S$ matrix,
 are represented by the dots. They are divided into 
the bound, decaying, capturing,  and  antibound states
(see, e.g.,  Refs.~\cite{[Ber68],[Ver87],[Lin93]}). The relation
(\ref{Berggren_comp}) involves 
the bound and decaying states and the contour $L_{+}$ lying
in the fourth quadrant
of the complex-$k$ plane.

In practical applications, one has to discretize the integral in
(\ref{Berggren_comp}) \cite{[Lio96],[Ver98]}:
\begin{equation}
\label{x9}
\int_{L_+} |u_k\rangle \langle\widetilde{u_k}| \, dk \simeq   
\sum_{i=1}^{N_d} |u_i\rangle \langle\widetilde{u_i}|,
\end{equation}
where $u_i(r) = \sqrt{\Delta_{k_i}} u_{k_i} (r)$ and 
$\Delta_{k_i}$ is the discretization step. It follows from the definition
of $u_i(r)$ that
\begin{eqnarray}
\label{x11}
\langle u_i | \widetilde{u_j}\rangle = \delta_{i,j},  
\end{eqnarray}
and the {\it discretized} Berggren 
relation (\ref{Berggren_comp}) takes the form:
\begin{eqnarray}
\sum_{n} |u_n\rangle \langle\widetilde{u_n}| +  
\sum_{i=1}^{N_d} |u_i\rangle \langle\widetilde{u_i}| \simeq 1.
\label{Berggren_comp_discr}  
\end{eqnarray}
This relation is formally identical to the standard completeness relation 
in a discrete basis and, in the same way, leads to the eigenvalue problem 
$H|\Psi\rangle = E|\Psi\rangle$. However, as the formalism of Gamow states is
non-hermitian, the matrix $H$ is complex symmetric.

Up to this point, the choice of the contour 
 in Eq.~(\protect\ref{Gamow_basis}) has been  completely arbitrary. In practice,
however, one wants to minimize the number of discretization points $N_d$  along
 $L_{+}$. This can be achieved  if the scattering functions on the contour
 (or, rather, their  phase shifts) 
change smoothly from point to point. This condition can be met if the contour 
does not lie in the vicinity of a pole, especially the narrow resonant state.
If this condition is met,  the states appearing in the
basis (\ref{Berggren_comp_discr})  can be naturally divided into:
\begin{itemize}
\item
{\it Bound states} -- lying on the imaginary $k$-axis (or negative real energy axis),
\item
{\it Narrow decaying states} -- lying close to
and below   the real $k$-axis (or below the positive
real energy axis). Those 
states
can be interpreted as {\it physical resonances of the system},
\item
{\it Non-resonant continuum} -- represented by the scattering states along
  $L_{+}$. Physically, those  are the building blocks of the non-resonant background.
\end{itemize}  
These definitions can be extended to   many-body states in the complex energy
(or momentum) plane. In the following, we shall clearly distinguish between
 resonant, resonance, and non-resonant  states.

The completeness relations derived above hold in every $(j,\ell)$ channel. Consequently,
in practical calculations, one has to take different contours 
for different  partial waves. As discussed below, the choice of the contour
depends on the distribution of resonant states in the complex $k$-plane.

In a number of papers (see, e.g., Refs. \cite{[Dus92],[For97],[Civ01]}),
 resonant states were applied to
problems involving continuum  in the  so-called {\it pole expansion},
  neglecting the contour integral in Eq.~(\ref{Berggren_comp}).
The importance of the contour contribution was investigated in 
Refs.~\cite{[Ver95],[Lin94],[Mic02],[Bet02],[Bet02a]}
 where it was concluded that 
if one is aiming at a detailed description, the non-resonant contribution must 
be accounted for.  This point will be clearly
seen in several examples discussed below.

\subsection{Completeness of the one-body Berggren basis: illustrative examples}
\label{lie_base_resonant_1corps}

In this section, we shall discuss examples of the Berggren completeness relation
in the one-body case. The s.p. basis is
generated by the spherical
Woods-Saxon (WS) potential:
\begin{mathletters}
\begin{eqnarray}
\label{WSpot}
V(r) &=& - V_0 f(r) - V_{\rm so} 4 \vec{l} \cdot \vec{s} \frac{1}{r}    
\frac{df(r)}{dr},\\
\label{wspot}
f(r) &=& \left[ 1+\exp \left({\frac{r-R_0}{d}}\right) \right]^{-1}.
\label{WSform}
\end{eqnarray}
\end{mathletters}
In all examples of this section, the WS potential has the radius $R_0$ = 5.3 fm,
diffuseness $d=0.65$ fm,
and the spin-orbit strength $V_{so}$ = 5.0 MeV.
The depth of the central part is  varied to simulate different situations. 

The complex contour corresponds to three straight segments in
the complex $k-$plane, joining the points: $k_0 = 0.0-i0.0$, 
$k_1=0.2-i0.2$, $k_2 = 0.5-i0.0$ and $k_3 = 2.0-i0.0$. The contour is
discretized with a different number of points: $n$=60, 80, 100, 120, 160, and
final results are obtained using the Richardson extrapolation method. 
In the examples considered in this section, we
shall expand the    $2p_{3/2}$ state, $|u_{\rm WS}\rangle$,
either weakly bound or resonant, in
the basis $|u_{\rm WS^B}(k)\rangle$ generated by the WS potential of a different depth:
\begin{eqnarray}\label{cexp}
|u_{\rm WS}\rangle &=& \sum_{i}c_{k_i}|u_{\rm WS^B}(k_i)\rangle \nonumber \\
 &+& \int_{L_+}  c(k) |u_{\rm WS^B}(k)\rangle\,dk,
\end{eqnarray}
cf. Eq.~(\ref{Berggren_comp}). In the above equation, the first term in the expansion
represents contributions from the resonant states while the second term 
is the non-resonant continuum contribution. 
Since the basis is properly normalized, the expansion amplitudes meet the
condition:
\begin{eqnarray}\label{cexpnorm}
\sum_{i}c^2_{k_i} + \int_{L_+}  c^2(k)\,dk = 1.
\end{eqnarray}
In all cases considered, 
the $0p_{3/2}$ and $1p_{3/2}$ orbitals  are well bound  
(by $\sim 40$ MeV and $\sim 18$ MeV, respectively) and do not play any role in the
 expansion studied.

\begin{figure}[htb]
  \begin{center}
 \leavevmode
  \epsfxsize=7.5cm
  \epsfbox{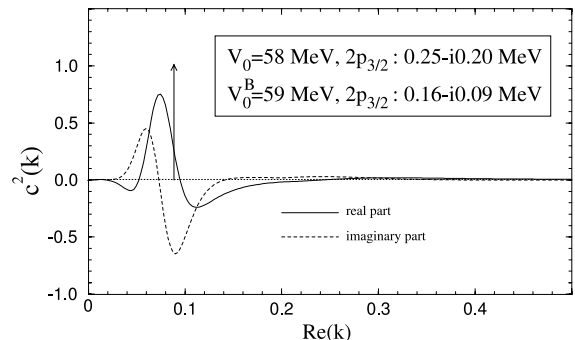}
\caption{Distribution of the squared amplitudes $c^2(k)$
of the s.p. state $2p_{3/2}$ of one WS potential ($V_0$ = 59 MeV) 
in the s.p. basis generated by another WS
potential ($V_0^{(B)}$=58 MeV). The amplitudes of both real (solid line)
and imaginary (dashed line) parts of the wave function
are plotted as a function
of Re[$k$]. The height of the arrow gives the squared amplitude
of the $2p_{3/2}$ resonance contained in the basis.
}
\label{usub}
\end{center}
\end{figure}

In the first example, we shall expand the $2p_{3/2}$ s.p. resonance (0.25--i0.20 MeV)
of a WS potential of the depth $V_0$=58 MeV in the basis generated by the WS 
potential of the depth  $V_0^{B}$=59 MeV (here the $2p_{3/2}$ s.p. resonance has
an energy of
0.16--i0.09 MeV). The density of the expansion 
amplitudes is shown in Fig.~\ref{usub}. One can see that the contribution
from the non-resonant
continuum is essential even though the $2p_{3/2}$ basis state 
is a resonance. In this example, as one might expect, the contribution from the
resonance state in the basis is dominant.

\begin{figure}[htb]
  \begin{center}
 \leavevmode
  \epsfxsize=7.5cm
  \epsfbox{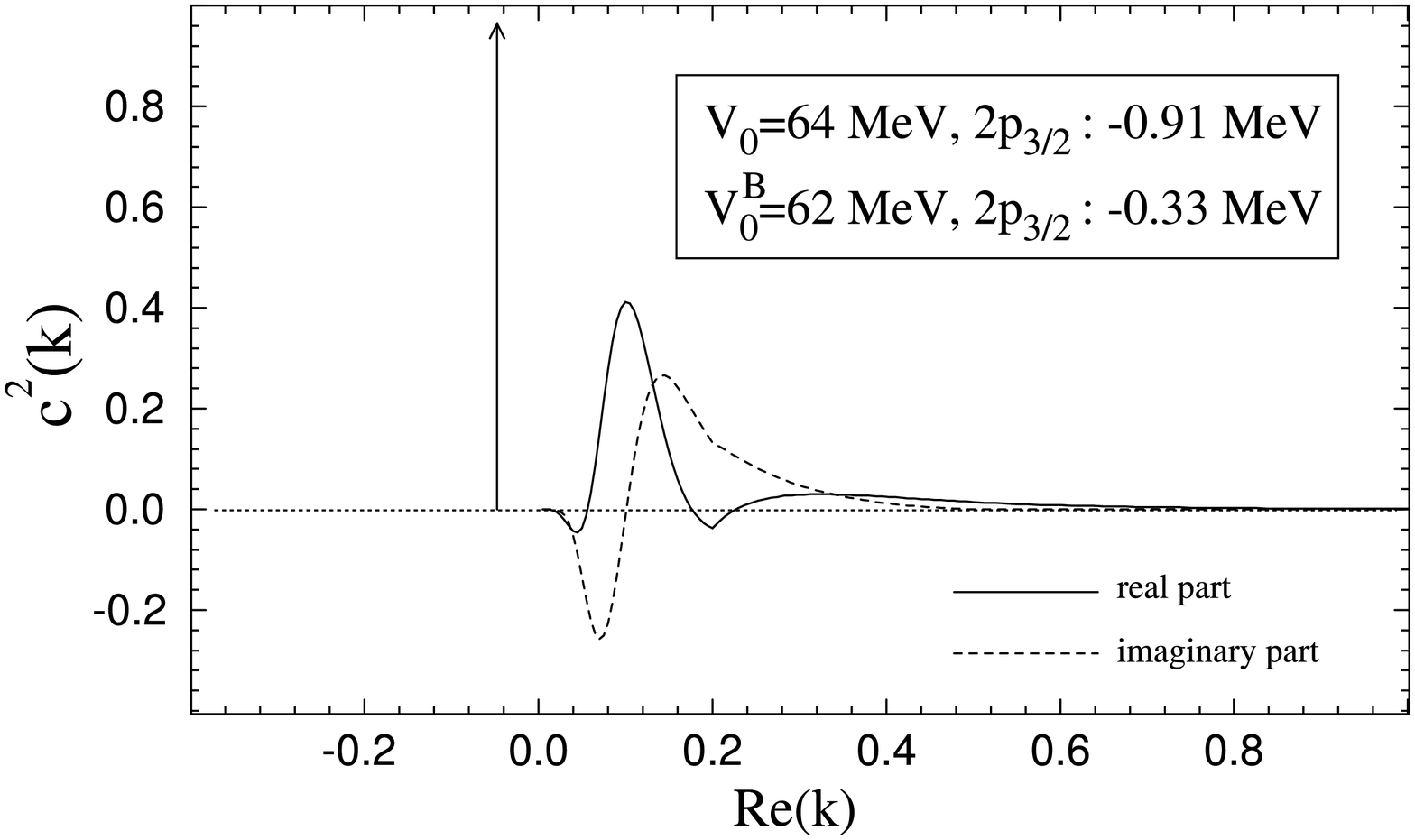}
\caption{Similar as in Fig. \protect\ref{usub} but for the bound
$2p_{3/2}$ s.p. state of the  WS potential with $V_0$=64 MeV
expended in the  basis generated by another WS
potential ($V_0^{(B)}$ = 62 MeV). The height of the arrow gives 
the squared amplitude  of the bound $2p_{3/2}$ state at the  value
of -Im[$k$] (the corresponding 
$k-$value is purely imaginary).
}
\label{bsbb}
\end{center}
\end{figure}

The second example shown in Fig. \ref{bsbb}
deals with the case of a  $2p_{3/2}$ state that
 is bound in both potentials. Here  $V_0$=64 MeV and 
$V_0^{(B)}$=62 MeV, and the  $2p_{3/2}$  state lies at $-$0.91 MeV  and $-$0.33 MeV,
respectively.
As in the previous case, the contribution from the resonant
(here: bound)  state in the Berggren basis
dominates and the contribution of the
non-resonant continuum is small although non-negligible. In this figure,
 one can  notice
 the small cusp at Re[$k$]=0.2, even though the density  $c^2(k)$ is an analytic function
of $k$. This apparent paradox is due to the fact that $c^2(k)$  is plotted  
as a function of Re[$k$] (Im[$k$]=0). Moreover,
the path in the complex plane is continuous but not derivable at 
$k$ = $0.2-i0.2$. These two aspects contribute to the appearance of the 
`discontinuous feature', which of course has no physical meaning.

\begin{figure}[htb]
  \begin{center}
 \leavevmode
  \epsfxsize=7.5cm
  \epsfbox{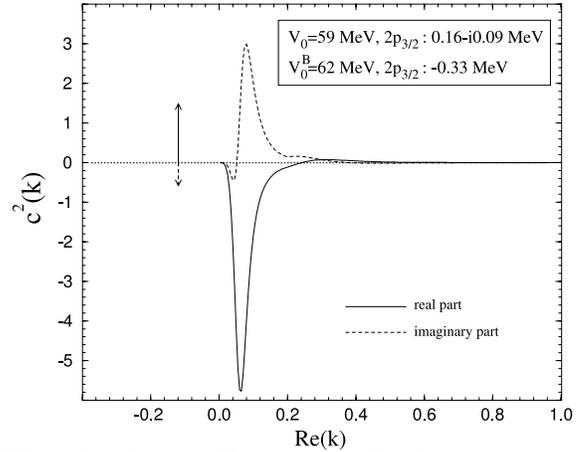}
\caption{Similar as in Fig. \protect\ref{usub} except for the 
$2p_{3/2}$ resonance of the  WS potential with $V_0$=59 MeV
expended in the  basis generated by another WS
potential ($V_0^{(B)}$ = 62 MeV).
}
\label{usbb}
\end{center}
\end{figure}

\begin{figure}[htb]
  \begin{center}
 \leavevmode
  \epsfxsize=7.5cm
  \epsfbox{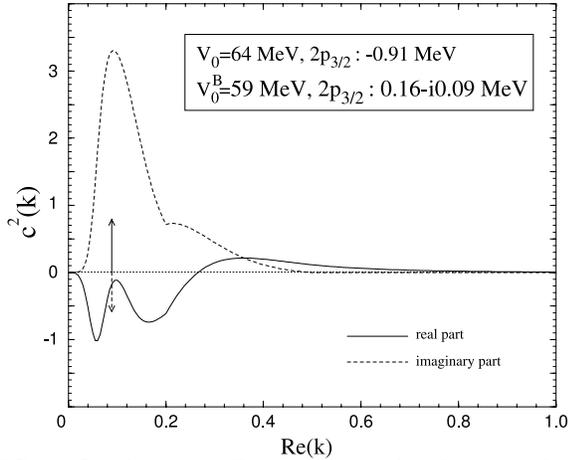}
\caption{Similar as in Fig. \protect\ref{usub} except for the 
$2p_{3/2}$ bound state of the  WS potential with $V_0$=64 MeV
expended in the  basis generated by another WS
potential ($V_0^{(B)}$=59 MeV) containing a resonance.
}
\label{bsub}
\end{center}
\end{figure}

\begin{figure}[htb]
  \begin{center}
 \leavevmode
  \epsfxsize=7.5cm
  \epsfbox{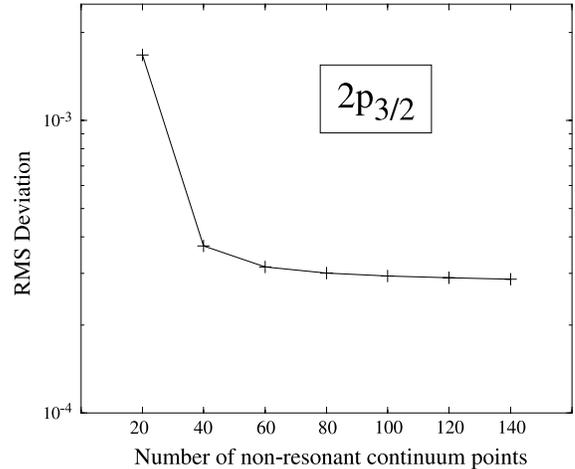}
\caption{The root mean square (RMS) deviation of the 
$2p_{3/2}$ wave function of the  WS potential ($V_0=64$ MeV)
obtained by a diagonalization in a basis generated by another WS potential 
($V_0^{B}$=59 MeV) from the exact wave function 
(obtained by a 
direct integration of the Schr\"odinger equation).
The RMS deviation is shown as 
a function of the number
of discretization points along the $2p_{3/2}$ contour.
}
\label{1body_prec}
\end{center}
\end{figure}

An interesting situation is presented in Fig.~\ref{usbb}. Here the unbound 
$2p_{3/2}$ state ($V_0$=59 MeV) is expanded in a WS basis containing
the bound $2p_{3/2}$ level ($V_0^{(B)}$=62 MeV).
Consequently,  the non-resonant continuum has to supply the imaginary part
of the resonance's wave function. 
The last example (Fig. \ref{bsub}) corresponds to $V_0$=64 MeV 
and  $V_0^{(B)}$=59 MeV. This is the most intriguing case since one expresses
a bound (real) state in the basis which contains only complex wave
functions (the contribution from well bound $0p_{3/2}$ and $1p_{3/2}$ s.p. 
states is negligible). In this case,  the non-resonant continuum 
annihilates  the imaginary component  of the $2p_{3/2}$ s.p. resonance 
contained in the
basis. Indeed, in this example, the contribution from the contour is dominant.
To see the convergence of the wave function obtained by the
expansion method, in Fig.~\ref{1body_prec} we show the 
root mean square
(RMS) deviation of the 
calculated $2p_{3/2}$ wave function from the exact result  as a function
of the number of discretization points along the contour. 
While the wave functions converge fairly quickly, the convergence
of complex energies  is slightly slower; hence,  one has to 
employ  the Richardson extrapolation method to get an energy precision of the
order of a keV.

\section{Gamow Shell Model Hamiltonian}\label{model}

The GSM  Hamiltonian applied in this work
consists of a one-body term and a zero-range
 two-body interaction. The spherical one-body potential was taken in a WS
form (\ref{WSpot}). 
In our study,
resonant  states are determined using the generalized shooting method for
bound states
which requires an exterior complex scaling.
The numerical algorithm for finding Gamow states for any finite-depth potential
has been tested on the example of the
P\H{o}schl-Teller-Ginocchio (PTG)
potential \cite{[Gin84]},  for which the resonance energies and
wave functions are known analytically. Energies of all PTG resonances with a
width of up to $90$ MeV are reproduced 
with a precision of at least $10^{-6}$ MeV.

Contrary to the traditional shell model, 
the effective interaction of CSM cannot be represented as 
a single matrix calculated for all nuclei in a given region. 
The GSM Hamiltonian contains  a real 
effective two-body force expressed in terms of space, spin, and isospin coordinates.
The matrix elements
involving continuum states are strongly system-dependent, and they have 
to be determined  for each case separately. This creates an additional difficulty,
but there is also a pay-off. Namely, the resulting two-body matrix elements
fully  take
into account the spatial extension of s.p. wave functions. 

In this work,
as a residual interaction we took the surface
delta interaction \cite{[Gre65]}
\begin{equation}
V(1,2) = -V_{\rm SDI} \delta\left(\vec{r}_1-\vec{r}_2 \right)\delta(r_1-R)
\end{equation}
with the the same value of $R$ as in the WS  potential.
In our   exploratory GSM calculations, we consider two cases:
(i) the chain of oxygen isotopes  with the inert $^{16}$O core 
and active 
neutrons  in the $sd$ shell,
and (ii) 
the helium chain 
 with the inert
$^{4}$He core and active neutrons  in the $p$ shell. 
The parameters of the GSM Hamiltonian are summarized in Table~\ref{GSMparameters}.
The WS potentials have been adjusted to s.p. states in one-neutron nuclei
$^{17}$O and $^5$He. 
For $^{17}$O,
the resulting WS   
$0d_{5/2}$ and $1s_{1/2}$ states are bound, with s.p.\ energies -4.142\,MeV 
and -3.272\,MeV, respectively, and
$0d_{3/2}$ is a resonance with the s.p.\ energy 0.898--i0.485\,MeV.
 The agreement with  experimental data 
($e_{5/2_1^+}^{exp}$=--4.143 MeV, $e_{1/2_1^+}^{exp}$ = -3.273 MeV, and 
$e_{3/2_1^+}^{exp}$ =  0.942 MeV, $\gamma_{3/2_1^+}^{exp}$ = 96 keV) is
excellent.  The strength of the SDI 
has been  adjusted for a given configuration space 
to the experimental two-neutron separation energy 
of  $^{18}$O.
Since only $0d_{3/2}$ is a s.p. resonance, 
we shall include only a $d_{3/2}$ non-resonant continuum.
In fact, the completeness relation requires taking 
the  non-resonant continua corresponding to all partial waves ($\ell,j$). 
However, if for a given partial wave no resonances are included in the basis,
the corresponding 
 non-resonant continua can be chosen
along the real momentum axis. 
Since, to the first order, the inclusion of these
 continua should only result in the renormalization of the effective interaction,
they can be  ignored in most cases, except for Sec.~\ref{fondam_O18}.
%
%
\begin{table}[tp]
\begin{center}
\begin{tabular}{c|ddddd}
    Variant            & $R$ (fm) & $d$ (fm) & $V_0$ (MeV) &  $V_{\rm so}$ (MeV) & $V_{\rm SDI}$ 
\\
      & (fm)   & (fm)  & (MeV) & (MeV)  & (MeV\,fm$^3$) \\
\hline
$^{17}$O      &  3.05 &  0.65  & 55.8  & 6.06 & 700   \\
$^{5}$He      &  2.00  &   0.65   &   47.0  & 7.50 & 1670 \\
\end{tabular}
\end{center}
\caption{
Parameters of the GSM Hamiltonian used in the calculations for the oxygen isotopes
(``$^{17}$O" parameter set) and the helium isotopes (``$^{5}$He" parameter set):
WS radius $R$, WS diffuseness $d$, WS strength $V_0$, spin-orbit strength $V_{\rm so}$,
and strength of the residual SDI interaction $V_{\rm SDI}$.
}\label{GSMparameters}
\end{table}
%
%

The nucleus $^{5}$He,  with
one neutron in the $p$ shell,  is unstable with respect to the neutron emission. Indeed,
the $J^{\pi}=3/2_1^{-}$
ground state of $^{5}$He lies 890\,keV above the neutron emission
threshold and its neutron width is large, $\Gamma$=600\,keV. The first excited state, 
$1/2_1^{-}$, is a very broad resonance ($\Gamma$=4\,MeV) that lies
4.89\,MeV above the  threshold. Our WS potential yields
single-neutron resonances $p_{3/2}$ 
and $p_{1/2}$ at  $E$=0.745--i0.32 MeV  and $E$=2.130--i2.936 MeV, respectively.
In our model space we take resonances 
$0p_{3/2}$, $0p_{1/2}$, and the two associated complex continua $p_{3/2}$ and 
$p_{1/2}$.
The strength of the SDI  has been  adjusted 
for a given configuration space to the experimental two-neutron separation energy 
of  $^{6}$He. 

For the $N$-body problem, the Hamiltonian matrix contains one- and
two-body matrix elements. The one-body part corresponds to s.p. energies of
basis states and contributes only to diagonal matrix elements.
In general, the calculation of two-body matrix elements 
 is performed by splitting the radial
integral into 16 terms corresponding to all different possible asymptotic
conditions of s.p. wave functions. Then, each term is regularized
separately by an appropriate choice of angle of the external complex scaling, cf.
Sec.~\ref{prol_analy}.

\section {Many-body completeness relation with Gamow states}\label{completmb}

The discretized basis (\ref{Berggren_comp_discr}) can be a starting point for
establishing the completeness relation in the many-body case, in a full analogy
with the standard shell-model in a complete discrete basis, e.g., 
the harmonic oscillator basis. In this case one has:
\begin{eqnarray}
\sum_{n} |\Psi_n\rangle \langle\widetilde{\Psi_n}| \simeq 1,
\label{Berggren_comp_discr_N_corps}  
\end{eqnarray}
where the $N$-body Slater determinants
$|\Psi_n\rangle$ have the form  $|\phi_1\ldots \phi_N\rangle $,
where $|\phi_k\rangle$ are resonance (bound and decaying) and 
scattering (contour)  s.p. states. The approximate equality in 
(\ref{Berggren_comp_discr_N_corps}) is an obvious consequence of the continuum 
discretization, similarly as in (\ref{Berggren_comp_discr}).
Like in the case  of s.p. Gamow states, the normalization of the Gamow
vectors in the configuration space is given by the squares of shell-model amplitudes:
\begin{eqnarray}
\sum_n c_n^2 = 1,
\end{eqnarray}
and not to the squares of their absolute values.

In the particular case of  two-particle states, the completeness relation  reads:
\begin{equation}
\sum_{i_1,i_2} \ket{\phi_{i_1} \; \phi_{i_2}}_J \,
               {_J} \bra{\phi_{i_1} \; \phi_{i_2}} \simeq 1.
\label{eqmn}
\end{equation}                            
This relation can be used to calculate the two-body matrix elements.

\subsection {Determination of many-body bound and resonance states} 
\label{seek_eigenstates}

Before discussing  completeness relations in the many-body case, 
let us  describe  the method of selecting many-body resonances. 
In a standard shell model, one often uses the Lanczos method to
find the low-energy eigenstates (bound states) 
in very large configuration spaces. This popular
method is unfortunately useless for the determination of many-body
resonances because of a huge number (continuum) of surrounding 
many-body scattering states, many of them having lower energy than the
resonances. 
A practical solution to this problem  is the two-step procedure proposed in 
Ref.~\cite{[Mic02]}:
\begin{itemize}
\item In the first step, one performs the pole approximation, i.e., 
 the Hamiltonian is diagonalized in a smaller basis
consisting of s.p.  resonant states only.
Here, some  variant of
the Lanczos method can be applied. The diagonalization  yields the 
first-order approximation to many-body resonances $|\Psi_i\rangle^{(0)}$, where index
$i$ ($i=1,\dots ,N$) enumerates all eigenvectors 
in the restricted space.
These eigenvectors  serve as starting
vectors (pivots)  for the second 
step of the procedure.
\item In the second step, one 
includes couplings to non-resonant continuum states in 
the Lanczos subspace generated by  
$|\Psi_j\rangle^{(0)}$ ($j \in[1,\dots ,N]$).
\item
Finally, one searches among the  $M$ solutions 
$|\Psi_{j;k}\rangle$, ($k=1,\dots ,M$) 
for the eigenvector which has the largest overlap with $|\Psi_j\rangle^{(0)}$.
\end{itemize}
This procedure is a variant of the Davidson method. Obviously, in the search 
for bound states, both Lanczos and Davidson methods can be used.
 In the following, we shall show that the procedure outlined  above
 allows for an efficient determination of physical states within 
the set of all eigenvectors
of a given Lanczos subspace.

As a representative  example, let us consider the cases of 
$^{18}$O and $^{20}$O with the core of $^{16}$O, i.e., two- and four-particle
systems, respectively. Here we employ 
the complex $d_{3/2}$  contour 
corresponding to two segments defined  by the points:
 $k_1 = 0+i0$, $k_2 = 0.2-i0.05$,
and $k_3 = 0.4-i0.0$.  
The contour is discretized with 17 and 9 points for $^{18}$O and $^{20}$O,
respectively. From all the possible many-body
configurations, we only keep  the Slater
determinants with an energy less than  5 MeV and with a 
width smaller than 1.7 MeV.

\begin{figure}[htb]
  \begin{center}
 \leavevmode
  \epsfxsize=7.5cm
  \epsfbox{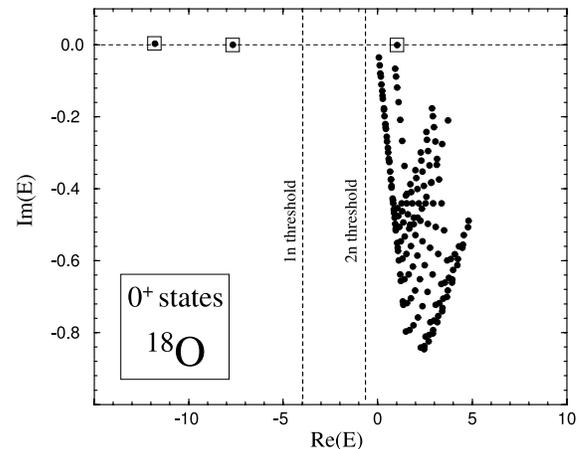}
\caption{Complex  energies of the $0^+$ states in $^{18}$O
resulting from the diagonalization of the GSM Hamiltonian. One- (1n) and two-neutron (2n)
emission thresholds are indicated. The physical bound and narrow resonance states
are marked by squares. The remaining eigenstates represent the non-resonant continuum.
}
\label{O18_plein_etats}
\end{center}
\end{figure}

\begin{figure}[htb]
  \begin{center}
 \leavevmode
  \epsfxsize=7.5cm
  \epsfbox{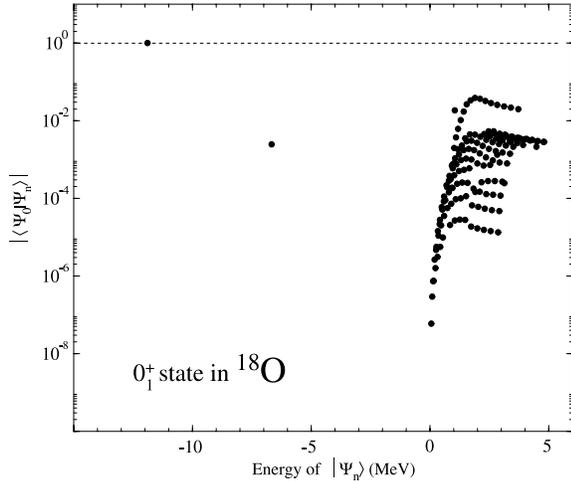}
\caption{Absolute value of the overlap between the ground state wave function 
$\Psi_0$ of $^{18}$O calculated without the coupling to the non-resonant 
continuum (pole approximation)
and the different $0_n^+$ eigenstates 
$\Psi_n$ of the GSM Hamiltonian
(calculated with the couplings to the $d_{3/2}$ non-resonant continuum), 
as a function of the total energy.
}
\label{overlap_O18}
\end{center}
\end{figure}

\begin{figure}[htb]
  \begin{center}
 \leavevmode
  \epsfxsize=7.5cm
  \epsfbox{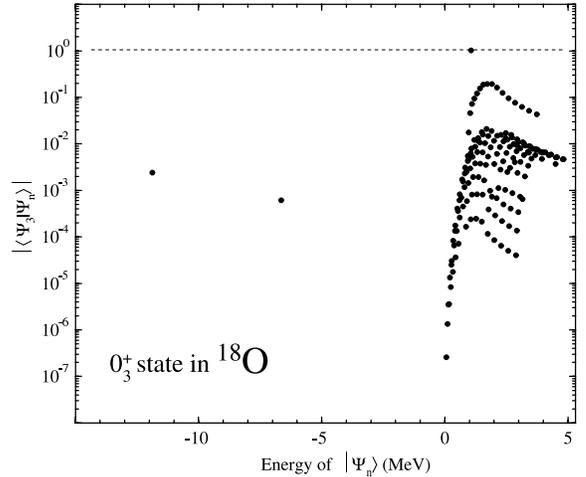}
\caption{Similar to Fig.~\protect\ref{overlap_O18} except  for
the $0^+_3$ state wave function of  $^{18}$O, $\Psi_3$.
}
\label{overlap_O18_3}
\end{center}
\end{figure}

The results of calculations
for the $0^+$ states in $^{18}$O are displayed in Fig. \ref{O18_plein_etats}.
In this case, one obtains
two bound states. An 
eigenstate lying just close to the real energy axis
just above the two-neutron threshold
is a candidate for a resonance. Even though
the  width of this state is very small, only
the overlap with the states calculated in the pole approximation 
can give the answer. Figure ~\ref{overlap_O18} shows
the overlap of the ground-state wave function $\Psi_0$ of $^{18}$O, 
 calculated in the pole approximation, with all the  $0^+$  eigenstates 
of the GSM Hamiltonian resulting from the full diagonalization.
 One can see
that only one state (the GSM ground state) has a significant overlap 
with $\Psi_0$;  hence, the identification of
the ground state wave function is unambiguous. Figure~\ref{overlap_O18_3}
illustrates  a more challenging case of the third 
$0^+_3$ state  of  $^{18}$O. In spite of the fact that this state is
embedded in the non-resonant $0^+$ continuum, its identification
is straightforward.
It is interesting to notice that those
 GSM  states in Fig. \ref{O18_plein_etats} that
 represent the non-resonant background tend to align along regular trajectories.
 As discussed in Refs. \cite{[Bet02],[Bet02a]}, the shapes of these trajectories directly
 reflect the geometry of the contour in the complex $k$-plane. In the 
 two-particle case, this information can be directly used to identify the resonance states.

\begin{figure}[htb]
  \begin{center}
 \leavevmode
  \epsfxsize=7.5cm
  \epsfbox{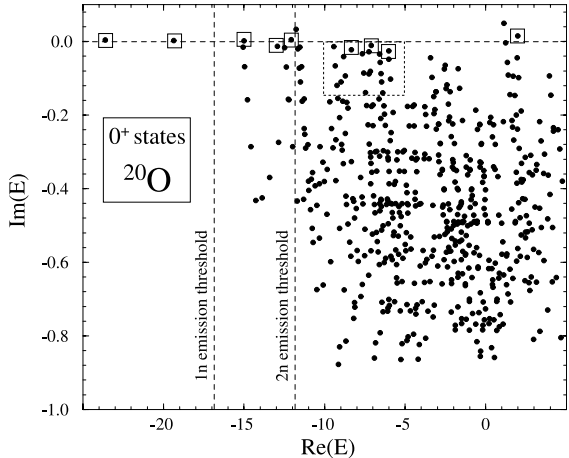}
\caption{Similar to Fig.~\protect\ref{O18_plein_etats} except for
the $0^+$ states in $^{20}$O. The small negative widths of several
unbound states are due to  the fact that  the number of
discretization points along the contour is fairly small (5) in this case.
}
\label{O20_plein_etats}
\end{center}
\end{figure}

Figure \ref{O20_plein_etats} shows the results of calculations
for the $0^+$ states in $^{20}$O. As compared to the 
$^{18}$O case, the number of many-body states is much larger
and the regular pattern of non-resonant states reflecting
the structure of the contour
is gone (the figure
represents the projection of four-dimensional trajectories onto two dimensional space).
While  the  two lowest (bound) 
states  can be simply identified by inspection,  
for the higher-lying states it is practically impossible to
separate the resonances
from the non-resonant continuum. However, 
the procedure outlined above makes it possible
 to identify unambiguously the many-body resonance states.
 On the other hand, the method proposed in Refs. \cite{[Bet02],[Bet02a]}
 cannot be easily applied.

\begin{figure}[htb]
  \begin{center}
 \leavevmode
  \epsfxsize=7.5cm
  \epsfbox{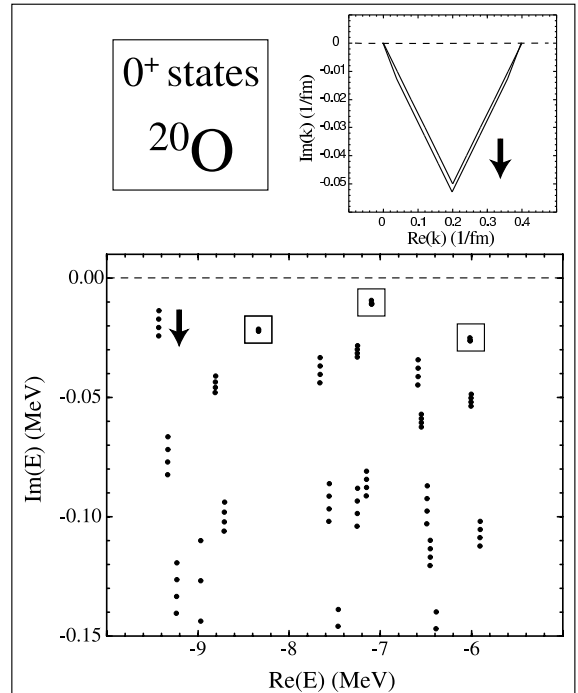}
\caption{The effect of small changes in the contour on the 
stability of resonant and non-resonant $0^+$ 
states in $^{20}$O. Top: the
contour in the complex-$k$ plane corresponding to  the $0d_{3/2}$
continuum. The direction of the contour's deformation
is indicated  by an arrow. The calculations were performed for four 
contours, each  divided into nine
segments (i.e., 10 discretization points); 
only the first  and maximally deformed contours are shown. Bottom:
the resulting shifts in positions of many-body states corresponding to
the complex energy region of 
 Fig.~\protect\ref{O20_plein_etats} marked by a dotted line. It is seen
 that the states identified as resonant are very stable with respect to small changes of
the  contour, while the states representing the non-resonant continuum move
 significantly in the direction indicated by an arrow.
}
\label{movecontour}
\end{center}
\end{figure}

The many-body resonances  should be stable with respect to small deformations of
 the contour (as the physical solutions should not be dependent on the deformation of
 the basis). This observation offers an independent criterion for identifying 
 resonance states. Figure~\ref{movecontour} shows 
the effect of a small   deformation of the contour on the 
stability of selected  $0^+$ 
states in $^{20}$O. As expected, only  the states
which have previously been  identified as resonances
 are  stable with respect to small changes of
the  contour; the states belonging to the non-resonant continuum 
 `walk'  in the complex energy plane following contour's motion.

\subsection{Completeness of the many-body Gamow basis: example of two
interacting particles}
\label{fondam_O18} \label{fondam_He6}

In this section, we shall discuss the completeness of the many-body basis
spanned by Gamow states. Since the number of configurations is growing extremely
fast with the number of valence nucleons, and this is  
enhanced by including  the non-resonant continuum, 
we shall restrict our discussion
to the case of two valence particles. By analyzing 
the behavior of the wave function as the number of basis states grows,
one can assess  the impact of various truncations  in the
valence space. This is especially important for calculations with  zero-range forces,
such as the  SDI  interaction employed in this work, 
which require  an energy cut-off. 
In the calculations contained in  this section, the configuration 
space consists of all the Slater determinants of energy less than 35 MeV. 
                                                  
As a first example, we shall consider the convergence of the ground-state
energy of $^{18}$O with the increasing size of the 
non-resonant phase space. 
The GSM  s.p. space consists of the $0d_{5/2}$, $1s_{1/2}$ orbitals and the
$0d_{3/2}$ Gamow resonance. This discrete basis is supplemented  by
adding successive continua: $s_{1/2}$, $p_{1/2}$, $p_{3/2}$,  \ldots , in the
decreasing order of their importance. Since  
$0d_{3/2}$ is a resonance,  a $d_{3/2}$ contour
should be complex, and we take it according to Sec.~\ref{seek_eigenstates}. 
Other non-resonant continua are real and for their path we choose  a straight segment with 
$0<k<1.3$~fm$^{-1}$ . The discretization of each continuum, whether real or complex, 
is made with 10 points.

\begin{figure}[htb]
  \begin{center}
 \leavevmode
  \epsfxsize=7.5cm
  \epsfbox{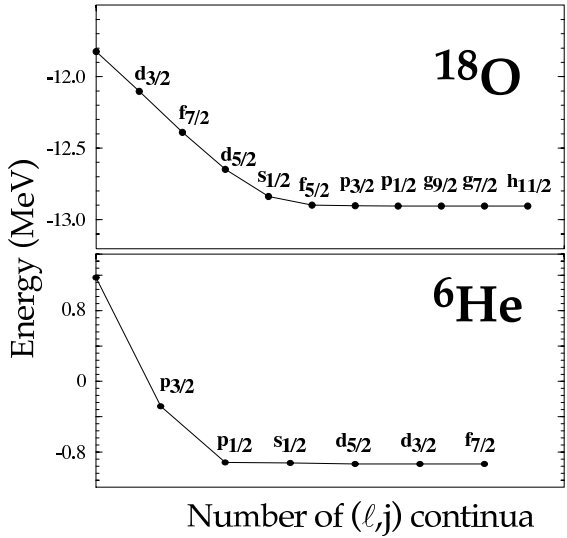}
\caption{
Ground-state energy of $^{18}$O (top) and $^{6}$He (bottom) 
as a function of the number of
different $(\ell,j)$ non-resonant continua in the valence space. 
For $^{18}$O  we employ 
the complex $d_{3/2}-$contour consisting of three straight segments connecting
the points: $k_0 = 0.0-i0.0$, $k_1 = 0.3-i0.2$, $k_2 = 0.5-i0$, and $k_3 = 1.3-i0$ (all in
fm$^{-1}$).   
}
\label{O18_convergence}\label{He6_convergence}
\end{center}
\end{figure}

As one can see in Fig.~\ref{O18_convergence} (top), the convergence is achieved with
the  $d_{3/2}$, $f_{7/2}$, $d_{5/2}$, $s_{1/2}$, and  $f_{5/2}$ contours; 
the contributions from all 
the remaining partial waves with $\ell>3$ and $\ell=1$ are practically negligible.
As a rule of thumb,
 the continua which contribute most to the binding energy 
are those which are associated either with  s.p. resonances in the basis 
or with  weakly bound s.p. states. This is the case with the $0d_{5/2}$, 
$1s_{1/2}$, and $0d_{3/2}$ orbitals, and also with the  $0f_{7/2}$ state
which has an energy of +7.43\,MeV and a width of 3.04\,MeV. 
Even though this wide resonance is absent in the basis, it
has nevertheless an indirect influence through the $f_{7/2}$ contour which
`keeps memory' of its presence. One may notice that the presence of the
(complex)  $d_{3/2}$ contour  affects not only the
real part of the ground-state energy but also its imaginary part; the 
ground-state energy has a spurious width of $\sim$ --130 keV in the pole approximation, 
and this width is reduced to  $\sim$ --2 keV when the
$d_{3/2}$ (10-point) contour is added. The spurious width of the ground state 
remains stable when other (real) contours are added.

%
%
\begin{table}[th]
\begin{center}
\begin{tabular}{ccc}
  Nucleus          & Configuration & $c^2$
\\
\hline
&  $0d^{2}_{5/2}$ & {0.872+i1.146 $\times 10^{-4}$} \\ 
&  $1s^{2}_{1/2}$ & {0.044--i5.973 $\times 10^{-6}$} \\ 
$^{18}$O&  $0d^{2}_{3/2}$ & {0.028--i6.624 $\times 10^{-3}$} \\ 
&  $L_{+}^{(1)}$ & {0.042+i4.709 $\times 10^{-3}$} \\ 
&  $L_{+}^{(2)}$ & {0.015+i1.795 $\times 10^{-3}$} \\ 
\hline
            & $0p^{2}_{3/2}$ & {0.891--i0.811} \\ 
            & $0p^{2}_{1/2}$ & {0.004--i0.079} \\ 
\rb{$^6$He} & $L_{+}^{(1)}$ & {0.255+i0.861} \\ 
            & $L_{+}^{(2)}$ & {$-$0.150+i0.029}              
\end{tabular}
\end{center}
\caption{
Squared amplitudes of different configurations in the
ground states  of two-neutron systems  $^{18}$O and $^6$He. The sum of squared amplitudes
of all Slater determinants, including $n$ particles in the non-resonant
continuum, is denoted by $L_+^{(n)}$.  See text for  details.
}
\label{table_GS_O18_complete}\label{table_GS_He6_complete}
\end{table}
%
%
The structure of the ground-state wave function of $^{18}$O calculated with the full
non-resonant continuum  discussed  in
the context of  Fig. \ref{O18_convergence} is given in Table 
\ref{table_GS_O18_complete}. One can see that the configuration with  2 neutrons
in the $0d_{5/2}$ shell dominates; the remaining configurations,
including those  with one and two neutrons in the non-resonant continuum,
 contribute 
with $\sim$ 15\%
to the wave function.   Imaginary parts
of squared amplitudes  are generally very  small. This is due to the
small width of the  $0d_{3/2}$ resonance included in the basis.

As a second example, we shall investigate  the energy convergence for the weakly bound 
ground state  of $^{6}$He. We shall assume that the structure of $^{6}$He
can be described in the full  $0p$ shell with two valence neutrons. 
Since the GSM  s.p. space consists of the $0p_{3/2}$, $0p_{1/2}$ Gamow resonances,
 the associated $p_{3/2}$ and $p_{1/2}$ non-resonant continua should be complex.
 Here, the
$p_{3/2}$ contour consists of three straight segments connected at points: 
$k_0 = 0.0-i0.0$, $k_1 = 0.3-i0.35$, $k_2 = 0.5-i0$, and $k_3 = 0.7-i0$. 
In the $p_{1/2}$ non-resonant channel, straight lines join the points: 
$k_0 = 0.0-i0.0$, $k_1 = 0.4-i0.5$, $k_2 = 0.5-i0$, and $k_3 = 0.7-i0$.    
 For all other
contours  we take a segment [0:0.7]~fm$^{-1}$
 of the real $k-$axis. The $p_{3/2}$  
contour is discretized with 20 points. All other contours, including the
complex $p_{1/2}-$contour, are discretized with 10 points.

As seen in Fig. \ref{He6_convergence} (bottom), the full energy convergence
  is attained 
with $p_{3/2}$ and $p_{1/2}$ contours;  other scattering waves with $\ell>1$ are
negligible. Coupling to the non-resonant $p_{3/2}$ and $p_{1/2}$ continua changes
not only  the ground-state energy but also its (spurious) width. In the pole
approximation, the calculated 
 ground state  has a huge  width of $\sim$ --2 MeV, which is reduced
to $\sim$ --650 keV when the  $p_{3/2}$ contour is added, and it reaches 
$\Gamma$$\sim$--10 keV when both $p_{3/2}$ and $p_{1/2}$ non-resonant
continua are included. 
Contrary to the case of $^{18}$O, the pole approximation
is totally unreliable, and it does not  even give a  rough 
approximation to the energy and wave function of the ground state of $^{6}$He. The main 
reason is that in this case
one attempts  to describe the bound state using the basis which does
not contain any bound state and, therefore, the non-resonant continuum
is  essential for compensating   the resonance contribution.
The analogous situation, discussed in the context of  a one-body problem, can be found
in Sect. \ref{lie_base_resonant_1corps} (Fig. \ref{bsub}). 

The structure of the ground-state wave function of $^{6}$He 
including all the ($\ell,j$) continua shown in Fig. \ref{He6_convergence}, is shown 
in Table \ref{table_GS_He6_complete}. One can see that
the configuration with two neutrons in the $0p_{3/2}$ shell dominates,
though the imaginary part  of the corresponding
squared amplitude is almost equal in magnitude to the real part. The
amplitude  of the  $0p^2_{1/2}$  configuration is small, whereas the 
 contributions from one  and two particles in the non-resonant
continuum, $L_+^{(1)}$ and $L_+^{(2)}$, are almost
equally important.

\section{GSM Study of oxygen isotopes} 
\label{chaine_O}

In this section we shall discuss the GSM results for a chain of oxygen
isotopes with several  valence neutrons ($N_{val}\geq 2$). In the calculations, we
assume the core of $^{16}$O. As discussed in Sec.~\ref{model}, the 
valence neutrons are distributed 
over the $1s_{1/2}$ and  $0d_{5/2}$ bound shells, the $0d_{3/2}$ Gamow resonance,  
and  the  discretized $d_{3/2}$ non-resonant continuum.
The complex $d_{3/2}$  contour 
corresponds to two segments defined  by the points:
 $k_1 = 0+i0$, $k_2 = 0.2-i0.05$,
and $k_3 = 0.4-i0.0$. It is discretized with 9 points. Consequently,  the
discretized s.p. GSM space (\ref{Berggren_comp_discr}) consists
of a total of  12 subshells on which valence neutrons are distributed.
Let us reiterate  that we also introduce  the cut-off
in the configuration space of the GSM. In this study, we shall include all
Slater determinants with an energy (width) smaller than 
5 MeV (1.7 MeV). Moreover, we shall take only Slater determinants with,
 at most,  2 neutrons in
the non-resonant continuum. We have checked that configurations  with
a larger number of neutrons in the non-resonant continuum are
of minor importance in all nuclei from $^{18}$O to $^{22}$O,
and their contribution is $<$ 0.01\% in all cases studied.
 Our aim is not to give the
 precise description of actual nuclei  (for this, one would need
 a realistic Hamiltonian and a larger configuration space), but rather
 to illustrate the method, its basic ingredients, and underlying features.

\subsubsection{Ground states of oxygen isotopes}

According to our calculations (see, e.g., Table~\ref{tables_GS_O18} for $^{18}$O),
the ground-state wave functions of $^{18-22}$O  are dominated by the single
shell model configuration $0d^{N_{\rm val}}_{5/2}$
 having a weight of 80-95\%; the non-resonant continuum
contribution is relatively small. A one-neutron continuum provides
between 0.1 and 1\% of the wave function, whereas a two-neutron continuum yields
a contribution which varies from $\sim 0.01$ to $\sim 0.1$\% in these nuclei. 
This means that the main effect of the non-resonant continuum in these states 
is to cancel a spurious
width induced by the  $0d_{3/2}$ resonance included in the basis,
 and their influence on
the real part of the energy can be neglected. 
%
%
\begin{table}[tp]
\begin{center}
\begin{tabular}{cc}
 Configuration & $c^2$  \\
\hline
$0d^{2}_{5/2}$ & {0.935+i1.836$\times 10^{-4}$} \\ 
$1s^{2}_{1/2}$ & {0.040--i1.560$\times 10^{-5}$} \\ 
$0d^{2}_{3/2}$  & {0.021--i4.973$\times 10^{-3}$} \\ 
$L_{+}^{(1)}$   & {3.913$\times 10^{-3}$+i4.412$\times 10^{-3}$} \\ 
$L_{+}^{(2)}$   & {$-$1.002$\times 10^{-4}$+i3.939$\times 10^{-4}$}             
\end{tabular}
\end{center}
\caption{Same as in Table \protect\ref{table_GS_O18_complete} 
but  only including the $d_{3/2}$ contour.
}
\label{tables_GS_O18}
\end{table}
%
%
It is interesting to compare   the squared
amplitudes  for $^{18}$O
  in Tables  \ref{table_GS_O18_complete}
and \ref{tables_GS_O18}. The values in Table  \ref{table_GS_O18_complete} have been
obtained by taking all the contours from $s_{1/2}$ to $h_{11/2}$;
whereas the results in
Table \ref{tables_GS_O18} have been calculated with only a
$d_{3/2}$ contour. Increasing the non-resonant space 
gives rise to   the depletion of the occupation  of  bound shells.

\begin{figure}[htb]
  \begin{center}
 \leavevmode
  \epsfxsize=7.5cm
  \epsfbox{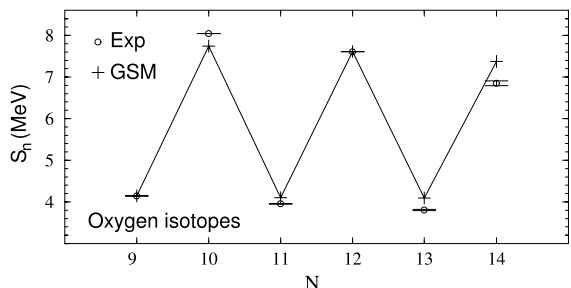}
\caption{
Calculated (crosses connected by a solid line) 
and experimental (circles with error bars) one-neutron separation energies in the 
$^{18-22}$O chain.  
}
\label{S1n_O_GSM}
\end{center}
\end{figure}

The predicted one-neutron separation energies $S_n$ for oxygen isotopes are 
displayed in Fig.~\ref{S1n_O_GSM} and
compared  with the data. The calculations reproduce experimental separation
energies, including the magnitude of the odd-even staggering.
The difference between GSM  and the data are in the worst case $\sim 500$ keV. 

\subsubsection{Level schemes  of oxygen isotopes}

\begin{figure}[htb]
  \begin{center}
 \leavevmode
  \epsfxsize=7.5cm
  \epsfbox{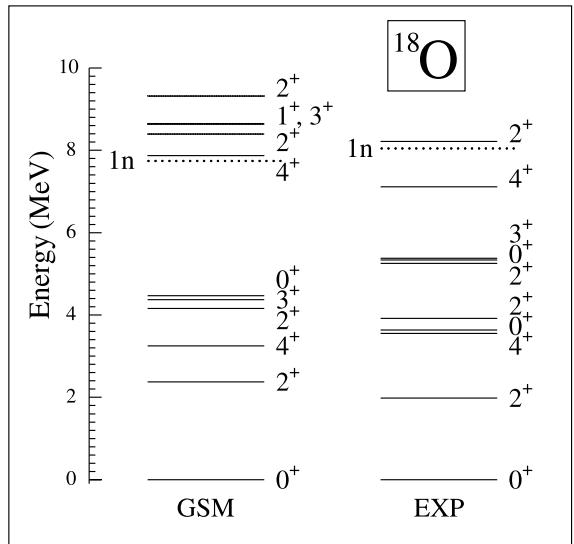}
\caption{
The GSM  level scheme of $^{18}$O 
calculated in the full $sd$ space of Gamow states and employing
the discretized (10 points) $d_{3/2}$ non-resonant continuum. 
The dashed lines indicate 
experimental and calculated one-neutron emission thresholds.
Experimental data are taken from Ref.~\protect\cite{[Fir96]}.
}
\label{spectre_O18}
\end{center}
\end{figure}

\begin{figure}[htb]
  \begin{center}
 \leavevmode
  \epsfxsize=7.5cm
  \epsfbox{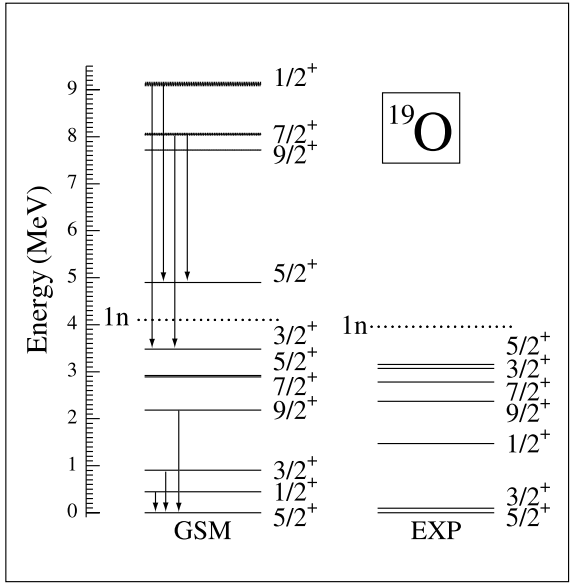}
\caption{
Same as in Fig. \protect\ref{spectre_O18} except for   $^{19}$O. 
  As the number of states becomes large above the one-neutron emission threshold,
  only selected resonances are shown.
The electromagnetic transitions listed in Table~\protect\ref{table_BEM_O19}
are indicated by arrows.
}
\label{spectre_O19}
\end{center}
\end{figure}

\begin{figure}[htb]
  \begin{center}
 \leavevmode
  \epsfxsize=7.5cm
  \epsfbox{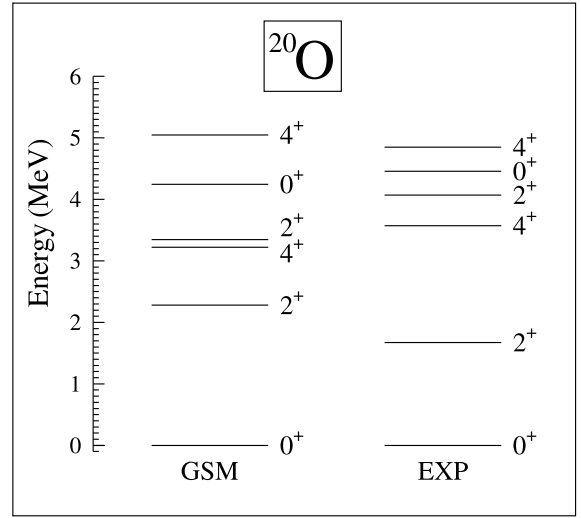}
\caption{
Same as in Fig. \protect\ref{spectre_O18} except for   $^{20}$O.
}
\label{spectre_O20}
\end{center}
\end{figure}

Shown in  Figs. \ref{spectre_O18} - \ref{spectre_O20} are  calculated and experimental
spectra of $^{18-20}$O. Let us stress that
the main purpose  of this comparison 
is to demonstrate the internal coherence of GSM results 
and the numerical feasibility of GSM in realistic applications. Nevertheless,
in spite of
the simplicity of the Hamiltonian, the overall
agreement between the GSM and experimental spectra is quite reasonable.

Figure \ref{spectre_O18} illustrates the case of $^{18}$O.
One can see
that the first three lowest states  
are relatively well described. Moreover, the one-neutron emission
threshold is  reproduced. In GSM, 
this threshold is obtained from the
difference of calculated ground-state energies of
$^{18}$O and $^{17}$O. It is important to note that all states above the
one-neutron emission threshold {\it are predicted to be resonances}, 
{i.e.}, their widths are positive, whereas all
states below this threshold are calculated to have small spurious negative widths. This
 feature  is certainly  not imposed on the model and it proves
both an internal consistency of the GSM as well as a correct numerical 
account of the completeness for many-body Gamow states. 
One should notice that in the calculations with a discretized continuum, it
is impossible to assure that the bound states have a width 
{\it exactly} equal to zero. Instead, one expects that this numerical artifact of
continuum discretization  becomes numerically unimportant 
with increasing the number of points on the  contour
(see Table~1 of \cite{[Mic02]}). With 10-point discretization, the
spurious width remains $\sim -5$ keV for all bound states of $^{18}$O, and
we consider this
a reasonable precision.
Similarly, for resonances lying very close to the lowest particle emission
threshold, hence having narrow widths, it may happen that their
width results  from the calculations as negative 
if the many-body completeness relation is numerically not well satisfied
(see  Fig.~\ref{O20_plein_etats} and related discussion).

The calculated level scheme  of $^{19}$O (see Fig. \ref{spectre_O19}) reproduces 
the main 
experimental features. The ground state is $5/2^{+}$, as observed,
but the higher-lying states $3/2^{+}_{1}$ and $1/2^{+}_{1}$  are given in the reversed
order. Nevertheless, the next four states are rather well reproduced, in spite of
the inversion of the $5/2^{+}_{2}$ and $3/2^{+}_{2}$ levels.
Finally, the GSM prediction for $^{20}$O is shown in  Fig.~\ref{spectre_O20}.
Here, the overall agreement between calculations and experiment is
best for all the isotopes studied. 

\subsubsection{Distribution of valence particles}\label{dedsdist}

Radial features of the density distribution of valence nucleons 
are always of great theoretical interest. They determine the nuclear size, electromagnetic transition rates,
polarization charges, and many other nuclear properties.
To assess the radial extent of the nucleonic distribution obtained in the GSM,
we investigate the monopole form factor defined through the normalized radial
Gamow wave functions $u_i(r)$ (\ref{Gamow_norm}):
\begin{equation}
{\rho}(r) = \sum_{i} n_i |u_i(r)\rangle \langle \tilde{u}_i(r)|,
\label{rho_def}
\end{equation}
where $n_i\equiv \langle a^+_na_n\rangle$ is the GSM occupation coefficient of the 
s.p. Gamow
orbital $i$. By definition, the form factor is normalized to the number of valence particles:
\begin{equation}
\int_{r=0}^{\infty'} {\rho}(r)\,dr = N_{\rm val},
\label{rho_norm}
\end{equation}
where the ``prime" sign indicates that the integral is calculated by means of complex scaling.

\begin{figure}[htb]
  \begin{center}
 \leavevmode
  \epsfxsize=7.5cm
  \epsfbox{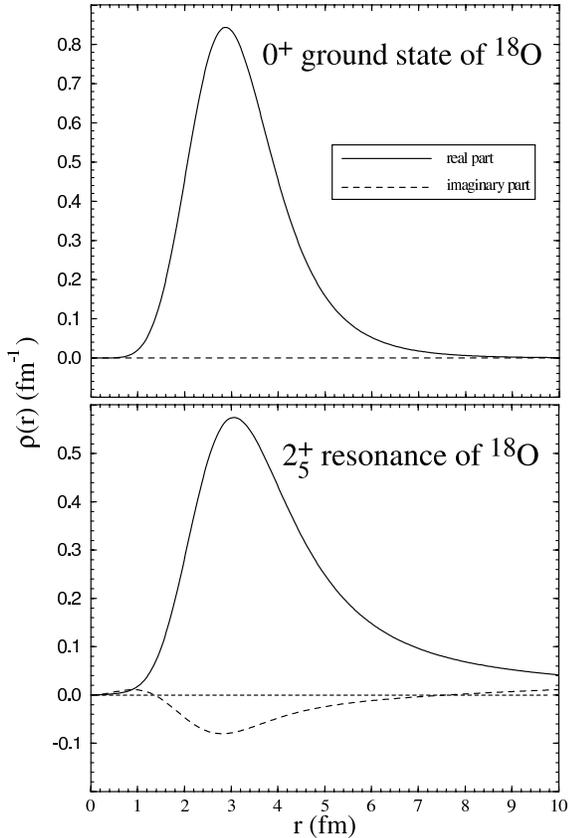}
\caption{
Monopole form factor of valence neutrons (\protect\ref{rho_def})
in the ground state  (top) and the $2^+_5$ resonance (bottom)
of $^{18}$O.
Real and imaginary parts of the density are shown with solid and dotted lines,
respectively. According to definition, Eq.~(\protect\ref{rho_norm}), the form factor is expressed in 
units of fm$^{-1}$.
}
\label{O18_densite}
\end{center}
\end{figure}

Figure~\ref{O18_densite}  shows the 
neutron form factors  for the ground state and  the
highly excited  resonance state  
$2^{+}_{5}$ of $^{18}$O.
As for all observables associated with bound states, the 
neutron form factor in the ground state  of $^{18}$O 
 is real even though the $0d_{3/2}$ resonant state is included
in the basis. The imaginary part of $\rho(r)$ is 
very small; its value is less
than $3 \times 10^{-5}$ for all values of $r$. On the other hand, 
for  the $2^{+}_{5}$ resonance,  having  a  width 
of $\sim$150 keV, 
the form factor is always complex. This feature is not an 
artifact of the discretization procedure. In fact,  the
imaginary part of the form factor has an interesting physical interpretation.
Namely, it
is associated with the uncertainty in the determination of the mean value
(given by the real part) \cite{[Ber68],[Gya72],[Ber96a],[Bol96],[Civ99]}. 
In other words, in the decaying state
the particle distribution cannot be given with an unlimited precision.

\subsubsection{Electromagnetic transition probabilities}

In the shell-model calculations with Gamow states, only radial 
matrix elements are treated differently as compared to the standard shell model.
This means  that the 
 electromagnetic (EM) transition
selection rules and the angular momentum and isospin algebra  do not change.
To calculate the EM transitions, one can no longer
use the long wavelength approximation because of the presence of
the non resonant
continuum. 
Indeed, for
the diagonal EM  matrix elements
$\langle n_i l_i j_i k_i|O| n_f l_f j_f k_f \rangle$ 
 ($k_i$ = $k_f$) between the   scattering states, 
 the complex scaling cannot be carried out  (see
discussion around Eq.~(\ref{delta_discret})).
 Furthermore, since in the long wavelength approximation
the  EM  operators behave like $r^\lambda$, 
one has to deal with   derivatives of delta functions, which is 
difficult to handle.
Without the long wavelength approximation, however, these matrix elements
become finite, because it is always possible to carry out  a complex scaling
with the Bessel function of the photon $j_L(qr)$, as $q \neq 0$.
Moreover, as they represent a
set of measure zero,  the diagonal non-resonant EM matrix elements 
 can be put to zero in the discretized calculation. As all
the other matrix elements can be regularized,
the EM  matrix elements are all well defined.

 Tables \ref{table_BEM_O18} and \ref{table_BEM_O19} display
the selected
EM transition rates  in  $^{18}$O and  $^{19}$O 
calculated in GSM  with and without the contribution from the 
non-resonant $d_{3/2}$ continuum.
In all calculations, we have taken the  effective neutron  charge  $\delta e_n = 0.5e$. 
While the radial form factor discussed in Sec.~(\ref{dedsdist}) determines 
the structure of monopole operators, the EM probabilities
probe the off-diagonal matrix elements; hence
the continuum coupling manifests itself differently.
As seen in Tables \ref{table_BEM_O18} and \ref{table_BEM_O19},
the non-resonant continuum plays an important  role both 
for transitions involving bound states only and for transitions involving
an  unbound state or states. In all cases,
 the real part of the transition rate is
reasonably approximated in the pole approximation.
 
 Similarly, as in the problem of spurious negative widths of 
bound states, the imaginary part of EM transitions between
bound states  must
disappear in the limit of the non-discretized continuum. 
Indeed, in our calculations
the imaginary part of such  a transition rate tends to zero when the coupling to the
non-resonant continuum is added.
In the  cases studied, taking only 10  points on the
$d_{3/2}$  contour
decreases the imaginary part by one to two orders of magnitude, depending on
the transition.  
The real part of the
transition probabilities changes by up to $\sim$ 10\% when 
the contribution from the contour is considered. This is nevertheless 
an important change if one notices
that the configurations with one and two neutrons 
in the non-resonant continuum amount to $\sim$ 0.1\% of bound
wave functions in these nuclei. For most transitions, the experimental values are reproduced
within a factor of  3 by the GSM, and we find this agreement satisfactory in light of
a very simple Hamiltonian employed.
 
\begin{table}[htb]
\begin{center} 
\begin{tabular}{c|ccr} 
Transition & Without contour & With contour  & Exp. \\ \hline 
$2_{1}^{+} \stackrel{\mbox{\tiny E2}}{\longrightarrow} 0_{1}^{+}$ &
{2.75--i0.001} & {2.68--i0.011} & {3.32} \\
$4_{1}^{+} \stackrel{\mbox{\tiny E2}}{\longrightarrow} 2_{1}^{+}$ &
{1.99+i0.015} & {1.94--i0.003} & {1.19}  \\
$0_{2}^{+} \stackrel{\mbox{\tiny E2}}{\longrightarrow} 2_{1}^{+}$ &
{1.66+i0.007} & {1.48+i0.006} & {17}   \\
$2_{2}^{+} \stackrel{\mbox{\tiny M1}}{\longrightarrow} 2_{1}^{+}$ &
{0.04--i0.000} & {0.04--i0.000} & {0.14} \\
$2_{2}^{+} \stackrel{\mbox{\tiny E2}}{\longrightarrow} 0_{1}^{+}$ &
{0.19+i0.004} & {0.18--i0.001} & 1.3
\end{tabular}
\end{center}
\caption{Electromagnetic GSM rates (all in W.u.) for the selected transitions
in $^{18}$O calculated (i) in the pole approximation  without the complex-energy contour
 and (ii) with the  $d_{3/2}$  contour representing the non-resonant continuum. Experimental 
data \protect\cite{[Fir96]} are shown in the last column.
}
\label{table_BEM_O18}
\end{table}

Table~\ref{table_BEM_O19} displays transition rates in $^{19}$O, indicated
in Fig.~\ref{spectre_O19} by arrows. The three lowest transitions are between
bound states and the GSM calculations with contour predict a very small imaginary part (less than 0.001 W.u.).
The next four transitions in  Table~\ref{table_BEM_O19} involve the unbound $7/2_{2}^{+}$, $1/2_{2}^{+}$,
and $5/2_{3}^{+}$ levels lying above the one-neutron threshold, and the corresponding transition rates are,
 in general, complex. The effect is particularly pronounced for
 the $1/2_{2}^{+} \stackrel{\mbox{\tiny E2}}{\longrightarrow} 5/2_{3}^{+}$ transition between
unbound states. As mentioned in Sec.~\ref{dedsdist}, the imaginary part gives the
 uncertainty  of the average  value \cite{[Ber96a],[Bol96],[Civ99]}. In all cases, the real part of the matrix element is slightly
 influenced by the interference with the non-resonant background,
\begin{table}[htbp]
\begin{center}
\begin{tabular}{c|ccr} 
Transition & Without contour  & With contour  & Exp.  \\ \hline
$3/2_{1}^{+} \stackrel{\mbox{\tiny M1}}{\longrightarrow} 5/2_{1}^{+}$ &
{0.01--i0.002} & {0.01--i0.0} &{0.088} \\
$1/2_{1}^{+} \stackrel{\mbox{\tiny M1}}{\longrightarrow} 3/2_{1}^{+}$ &
{0.02--i0.003} & {0.02+i0.0} & {0.0093} \\
$1/2_{1}^{+} \stackrel{\mbox{\tiny E2}}{\longrightarrow} 5/2_{1}^{+}$ &
{3.07+i0.010} & {3.07+i0.003} & {0.58} \\
$9/2_{1}^{+} \stackrel{\mbox{\tiny E2}}{\longrightarrow} 5/2_{1}^{+}$ &
{1.78+i0.000} & {1.74--i0.006} & {$<$1} \\
$7/2_{3}^{+} \stackrel{\mbox{\tiny E2}}{\longrightarrow} 5/2_{3}^{+}$ &
{0.12--i0.062} & {0.13--i0.037} &   \\
$1/2_{3}^{+} \stackrel{\mbox{\tiny E2}}{\longrightarrow} 5/2_{3}^{+}$ &
{4.57+i0.430} & {4.58+i0.373} &  \\
$7/2_{3}^{+} \stackrel{\mbox{\tiny E2}}{\longrightarrow} 3/2_{2}^{+}$ &
{0.12+i0.038} & {0.13+i0.054} & \\
$1/2_{3}^{+} \stackrel{\mbox{\tiny E2}}{\longrightarrow} 3/2_{2}^{+}$ &
{0.82--0.034} & {0.85--i0.012} &
\end{tabular}
\end{center}
\caption{Same as in Table~\protect\ref{table_BEM_O18} except for 
electromagnetic transitions in $^{19}$O shown in  Fig.~\protect\ref{spectre_O19}. 
}
\label{table_BEM_O19}
\end{table}

\section{GSM study of helium isotopes} 
\label{chaine_He}

A description  of the
neutron-rich helium isotopes, including 
 Borromean nuclei $^{6,8}$He,  is a challenge for the GSM.
 $^{4}$He is a 
well-bound system with the one-neutron emission threshold at 20.58\,MeV. On the
contrary, as discussed in Sec.~\ref{model}, the nucleus $^{5}$He
is a broad resonance. The two-neutron system 
 $^{6}$He, on the contrary, is bound with 
the two-neutron emission threshold at 0.98\,MeV and one-neutron emission
threshold at 1.87\,MeV. The first excited state $2_1^+$ at 1.8\,MeV in 
$^{6}$He is neutron unstable with a width $\Gamma$=113\,keV. 

The s.p.  configuration  space  used includes both resonances 
$0p_{3/2}$, $0p_{1/2}$ and the two associated complex continua $p_{3/2}$ and 
$p_{1/2}$, which are discretized with 5 points each. 
The fact that the resonances in the basis are so broad requires  particular
attention when selecting points along the contour.
 They are given in Table \ref{contis_points}.
%
%
\begin{table}[tp]
\begin{center}
\begin{tabular}{ccc}
         point No. & $p_{3/2}$ & $p_{1/2}$ 
\\
\hline
 1  & 0.01$-$i0.01 & 0.05$-$i0.05 \\
 2  & 0.10$-$i0.06 & 0.20$-$i0.43 \\
 3  & 0.20$-$i0.095 & 0.40$-$i0.45 \\
 4  & 0.30$-$i0.06 & 0.60$-$i0.43 \\
 5  & 0.40$-$i0.01 & 0.75$-$i0.05 
\end{tabular}
\end{center}
\caption{
Discretized  $p_{3/2}$ and $p_{1/2}$ contours in the complex-$k$ plane (in
fm$^{-1}$) representing
the non-resonant continuum in the He calculation.
}
\label{contis_points}
\end{table}
%
%
In this case, we cannot significantly increase the 
density of points along the contour  because the calculation for the
heavier helium isotopes with more valence neutrons 
 would not be feasible. On the other hand, we do not
introduce any  restriction on energies and widths
 of Slater determinants included. Also, no restriction on the  number 
of neutrons in the non-resonant continuum is imposed, with the only exception 
being  $^{9}$He, where we allow for at most 4 neutrons to occupy ``contour shells". Contrary to
the discussion in Sect. \ref{fondam_He6}, we neglect all remaining
real continua in
the present calculation. We shall see that this will have an impact on the relative
weight of different configurations in the wave function.

\subsubsection{Spectra of helium isotopes}

\begin{figure}[htb]
  \begin{center}
 \leavevmode
  \epsfxsize=7.5cm
  \epsfbox{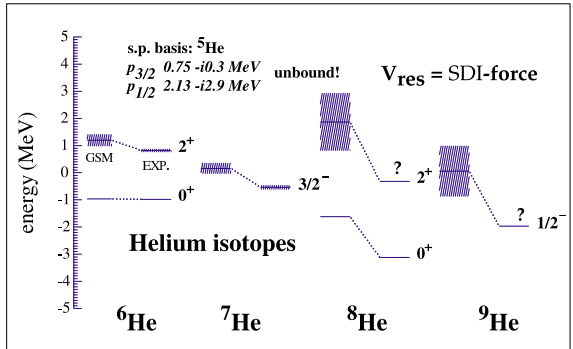}
\caption{
Experimental (EXP) and predicted (GSM)  binding energies of $^{6-9}$He
as well as energies of $J^\pi$=2$^+$ states in $^{6,8}$He.
The resonance widths are indicated by shading.
The energies are given with respect to the core of $^{4}$He.
}
\label{He_spectrum}
\end{center}
\end{figure}

The energies of the lowest GSM states of helium isotopes are
shown In Fig.~\ref{He_spectrum},  and the
structure of their ground state wave functions is given in
Table~\ref{Heliumsconf_GS}.
%
%
\begin{table}[htb]
\begin{center}
\begin{tabular}{ccc}
 Nucleus          & Configuration & $c^2$
\\
\hline
               &  $0p^{2}_{3/2}$    & 0.870--i0.736 \\
               &  $0p^{2}_{3/2}$    & 0.007--i0.076 \\
   \rb{$^6$He}             &  $L_{+}^{(1)}$     & 0.271 +i0.752 \\
               &  $L_{+}^{(2)}$     & $-$0.147+i0.060 \\
\hline
               &  $0p^{3}_{3/2}$           & {1.110--i0.879} \\ 
               &  $0p^{2}_{3/2} 0p_{1/2}$  & {0.006--i0.029} \\
               &  $0p_{3/2} 0p^{2}_{1/2}$  & {0.022--i0.042} \\
\rb{$^7$He}    &  $L_{+}^{(1)}$            & {0.050 +i0.951} \\ 
               &  $L_{+}^{(2)}$            & {$-$0.185+i0.008} \\ 
               &  $L_{+}^{(3)}$            & {$-$0.002--i0.009} \\
\hline
               &  $0p^{4}_{3/2}$    & {0.296--i1.323} \\ 
               &  $0p^{2}_{3/2} 0p^{2}_{1/2}$  & {$-$0.060--i0.158} \\ 
               &  $L_{+}^{(1)}$  & {1.596+i1.066} \\ 
\rb{$^8$He}    &  $L_{+}^{(2)}$  & {$-$0.728+i0.630} \\ 
               &  $L_{+}^{(3)}$  & {$-$0.125--i0.204} \\ 
               &  $L_{+}^{(4)}$  & {0.020--i0.012} \\ 
\hline
               &  $0p^{4}_{3/2} 0p_{1/2}$  & {0.180--i1.328} \\ 
               &  $L_{+}^{(1)}$  & {1.596+i0.734} \\ 
{$^9$He}       &  $L_{+}^{(2)}$  & {$-$0.584+i0.801} \\ 
               &  $L_{+}^{(3)}$  & {$-$0.217--i0.193} \\ 
               &  $L_{+}^{(4)}$  & {0.025--i0.034} 
\end{tabular}
\end{center}
\caption{
Squared amplitudes of different configurations in the
ground states  of $^{6-9}$He. 
The s.p. space consists of both 
$0p_{3/2}$, $0p_{1/2}$ Gamow resonances and the two associated complex continua $p_{3/2}$ and 
$p_{1/2}$ which are discretized with 5 points each. 
}
\label{Heliumsconf_GS}
\end{table}
%
%
As seen in  Table~\ref{Heliumsconf_GS},
the non-resonant
continuum contributions  $L_+^{(n)}$ are always essential, and, in some cases
(e.g., $^{8,9}$He),  they dominate the structure of the ground-state wave function.
Moreover, as can be seen in the example of $^8$He,
configurations with many neutrons in the non-resonant continuum are essential for 
fulfilling the
completeness relation. In this particular case, 
the $L_{+}^{(1)}$  contribution is
even more important than the contribution from the resonant states, and
even the $L_{+}^{(4)}$ 
configuration gives a non-negligible contribution of the order of 1-2\%. ($L_{+}^{(n)}$
amplitudes  seem to decrease with $n$.)
 Without the contour,  the predicted ground-state energy of $^{8}$He 
 is +2.08 MeV  and the  spurious width is huge, $\Gamma= -4.16$\,MeV. 
 The inclusion of scattering states  lowers the binding  energy to 
$-1.6$ MeV. Clearly,
 the pole approximation fails miserably in this case.

It is instructive  to compare the wave
function decomposition in Tables \ref{Heliumsconf_GS} (only $p_{3/2}$ and $p_{1/2}$ 
contours are included) and  \ref{table_GS_He6_complete} (all contours up to  $f_{7/2}$ 
are included). As
expected, the spread of the ground-state wave function over 
non-resonant continuum states 
is larger  in the latter case,  but this effect is relatively less important 
than in the case of $^{18}$O (cf.  Fig. \ref{He6_convergence}).

Our calculations reproduce the most important feature of $^6$He
and $^8$He: {\em the ground state
is particle-bound, despite the fact that all the basis states lie in the continuum.}
In spite of a very crude Hamiltonian,
rather limited configuration space, etc.,  the calculated
 ground state energies 
 shown in Fig.~\ref{He_spectrum}
  reproduce surprisingly well the
experimental data. The odd-$N$ isotopes of $^{7,9}$He are calculated to be neutron resonances.
The neutron separation energy anomaly (i.e.,
the {\em increase} of one-neutron separation energy
when going from $^6$He
to  $^8$He) is reproduced.
One can  see that $^{8}$He is predicted to be better bound  than
$^{6}$He, though, contrary to the data, the two-neutron separation energy of
$^{8}$He is smaller than that of $^{6}$He. 
Also, the energies of excited $2_1^{+}$ states  are in  fair agreement
with the data.

 The present calculations for helium isotopes have a systematic 
tendency to underbind.
The spurious width of the ground
state of $^{8}$He is $\sim$ 100 keV, largely
 due to  the   $0p_{3/2}$ resonance. For that reason, the 
 removal of the spurious width (which reflects the accuracy of our calculations)
cannot be done here as precisely  as in the  oxygen case. In order to check
the stability of the results for $^{6,8}$He,
we carried out additional calculations  increasing the number of points 
along the complex contour. We have 
found that the value of the two-neutron separation energy
difference, $S_{2n}(^{8}$He) -  $S_{2n}(^{6}$He), 
which is negative with 5 states in each non-resonant continuum,
 depends sensitively on
the number of discretization points. This is because
$S_{2n}(^{8}$He) increases fast with the size of the non-resonant phase space. While
with our schematic Hamiltonian we
were unable to find a fully satisfactory result for the ground-state energy of $^{8}$He 
 just by  increasing the number of discretization
points,  we can 
see that there is a systematic improvement  when the number of non-resonant
shells increases. Therefore, one would be  tempted to associate the so-called
`helium anomaly' in the position of two-neutron emission threshold 
when going from $^{6}$He to $^{8}$He
with the strong enhancement in  the occupation of non-resonant continuum states.

\subsubsection{Distribution of valence neutrons in $^{6}$He}

\begin{figure}[htb]
  \begin{center}
 \leavevmode
  \epsfxsize=7.5cm
  \epsfbox{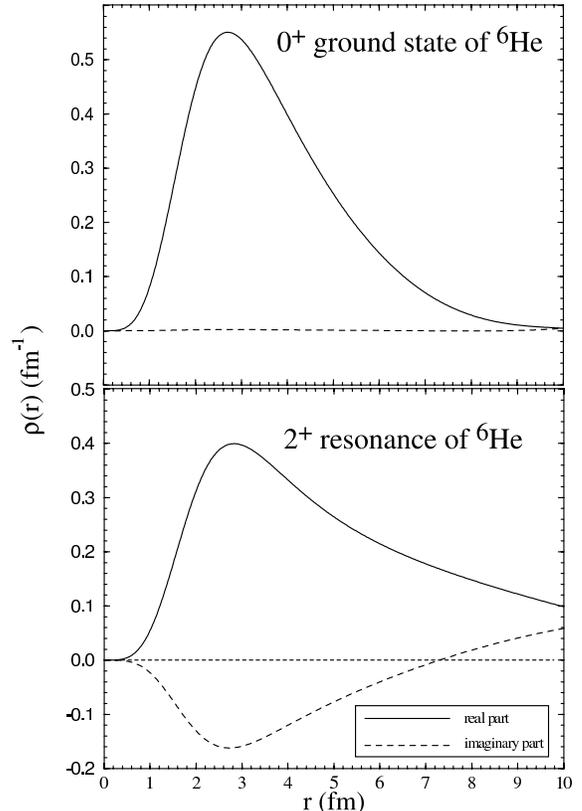}
\caption{Monopole   form factor of valence neutrons for the ground state  (top) 
and for the lowest $2^+_1$ resonance (bottom).of
$^{6}$He.
}
\label{He6_densite}
\end{center}
\end{figure}

Form factors of valence neutrons (\ref{rho_def}) in $0^{+}_{1}$ and 
$2^{+}_{1}$ states of $^{6}$He are shown in Figs.~\ref{He6_densite}.
In this calculation, we have taken
discretized $p_{1/2}$ and $p_{3/2}$ continua with 20 points each. Even with
this very high number of continuum states, the imaginary part of the form factor in the
ground state  of $^{6}$He is still
$\sim$ 2$\times 10^{-3}$\,fm$^{-1}$. 
It is interesting to see that the radial
distribution of neutrons in this case extends well above the range of
the one-body  potential, as expected for 
a weakly bound system. 
In the case of  the first excited (resonance)  $2^+_1$ state,
 $\rho(r)$ is complex; real and imaginary parts of the density are of a
comparable size, which is obviously related to the large width 
($\sim$ 500 keV) of this state.

\section{Summary}\label{summary}

This paper contains the detailed description of the Gamow Shell Model,
which can be viewed as a straightforward
extension of the standard diagonalization shell model that allows
for a consistent treatment 
of  bound states and the particle continuum, including both resonances and
the non-resonant background. Our first application,
based on a realistic, albeit simple,
 two-body SDI force, 
concerns  many-neutron  systems, including  states and nuclei near
the neutron-emission threshold.

In this work, we succeeded in overcoming several obstacles which
 traditionally plagued  previous
continuum shell model applications. 
In addition to the successful inclusion of
the continuum-continuum coupling by means of the 
complex rotation technique (exterior complex scaling),
we incorporated  the non-resonant
part of the continuum. This has been achieved by discretizing
the contour in the complex $k$-plane for each partial wave.
Another problem which has been solved in our study
is the isolation of  resonance  states. As a result of the GSM
diagonalization, one obtains a multitude of states corresponding to the
many-body continuum, some being  resonances and some representing the
non-resonant background. Our work offers a simple 
prescription on how to identify the resonance states from the multitude of 
complex-energy eigenstates of the GSM Hamiltonian.

The GSM Hamiltonian is composed of a one-body term and a two-body residual interaction
which is  directly written  in terms of space, spin, and isospin coordinates.
The two-body  matrix elements have 
to be determined   separately for each case by means of radial integration.
 As a result,  the  resulting  Hamiltonian matrix fully  takes
into account the continuum coupling,  in particular
the spatial extension of s.p. wave functions, determining  the physics of  halo nuclei.

We demonstrated that the
contribution from the non-resonant continuum is essential, especially
for unbound and  near-threshold states. In some cases
(e.g., $^{8,9}$He) non-resonant continuum components 
 dominate the structure of the g.s.
wave function. 
The so-called `pole approximation' (resonant
state expansion) breaks down in such cases. In addition, the inclusion of the
non-resonant continuum also impacts results for bound states.

The results of our first  calculations for binding energies, spectra,  electromagnetic
matrix elements, and nucleonic distributions  are 
very encouraging.  In particular, pairing correlations
due to the continuum-continuum scattering can  bind the ground states of
$^{6,8}$He with a completely unbound basis provided by the s.p. resonances
of $^5$He. The  `Helium anomaly'  (an {\it increase} in one- and two-neutron
separation energy when going from $^6$He to $^8$He) is explained in terms of the 
neutron  scattering to the non-resonant   background. 
In all cases considered, our  calculations yield neutron resonances  
above the  calculated neutron threshold -- a property that is not guaranteed {\em a priori}
by the formalism.

Other applications of GSM, including the case of open proton 
{\em and} neutron shells,   also employing more realistic  effective interactions,
are in progress. We are also working on 
the optimization of the non-resonant
part of calculations (choice of the contours, distribution of discretization points, etc.).
We are convinced that the Gamow Shell Model will become a very
 useful theoretical tool
unifying structure and reaction aspects of weakly bound nuclei.

%
%
\acknowledgements
This work was supported in part by the U.S.\ Department of Energy
under Contract Nos.\ DE-FG02-96ER40963 (University of Tennessee) and
 DE-AC05-00OR22725
with UT-Battelle, LLC (Oak Ridge National Laboratory).


\begin{thebibliography}{10}

\bibitem{[Dob97a]}
{J. Dobaczewski and W. Nazarewicz, Phil. Trans. R. Soc. Lond. A {\bf 356}, 2007
  (1998)}.

\bibitem{[Cau02]}
{E. Caurier, F. Nowacki, and A. Poves, Eur. Phys. J. A {\bf 15}, 145 (2002)}.

\bibitem{[Cau02a]}
{E. Caurier and G.Martinez-Pinedo, Nucl. Phys. {\bf A704}, 60c (2002)}.

\bibitem{[Ots02]}
{T. Otsuka, Y. Utsuno, R. Fujimoto, B.A. Brown, M. Honma, and T. Mizusaki, Eur.
  Phys. J. A {\bf 13}, 69 (2002); Eur. Phys. J. A {\bf 15}, 151 (2002)}.

\bibitem{[Cor02]}
{L. Coraggio, A. Covello, A. Gargano, N. Itaco, T.T.S. Kuo, D.R. Entem, and R.
  Machleidt, Phys. Rev. C {\bf 66}, 021303 (2002)}.

\bibitem{[Bro02]}
{B.A. Brown, Nucl. Phys. {\bf A704}, 11c (2002)}.

\bibitem{[Tho51]}
{R.G. Thomas, Phys. Rev. {\bf 81}, 148 (1951); {\bf 88}, 1109 (1952)}.

\bibitem{[Ehr51]}
{J.B. Ehrman, Phys. Rev. {\bf 81}, 412 (1951)}.

\bibitem{[Fes62]}
{H. Feshbach, Ann. Phys. (NY) {\bf 19}, 287 (1962)}.

\bibitem{[Fan61]}
{U. Fano, Phys. Rev. {\bf 124}, 1866 (1961)}.

\bibitem{[Mah69]}
{C. Mahaux and H. Weidenm\"{u}ller, {\it Shell Model Approaches to Nuclear
  Reactions} (North-Holland, Amsterdam, 1969)}.

\bibitem{[Bar77]}
{H.W. Bartz, I. Rotter, and J. H\"ohn, Nucl. Phys. {\bf A 275}, 111 (1977);
  {\it ibid.} {\bf A 307}, 285 (1977)}.

\bibitem{[Phi77]}
{R.J. Philpott, Nucl. Phys. {\bf A289},  109 (1977)}.

\bibitem{[Hal80]}
{D. Halderson and R.J. Philpott, Nucl. Phys. {\bf A345}, 141 (1980)}.

\bibitem{[Rot91]}
{I. Rotter, Rep. Prog. Phys. {\bf 54}, 635 (1991), and references quoted
  therein}.

\bibitem{[Oko03]}
{J. Oko{\l}owicz, M. P{\l}oszajczak, and I. Rotter,  Phys. Rep. {\bf 374},
  271 (2003)}.

\bibitem{[Ben99]}
{K. Bennaceur, F. Nowacki, J. Oko{\l}owicz, and M. P{\l}oszajczak, Nucl. Phys.
  {\bf A651}, 289 (1999)}.

\bibitem{[Ben00b]}
{K. Bennaceur, N. Michel, F. Nowacki, J. Oko{\l}owicz, and M.P{\l}oszajczak,
  Phys. Lett. {\bf 488B}, 75 (2000)}.

\bibitem{[Ben00c]}
{K. Bennaceur, F. Nowacki, J. Oko{\l}owicz, and M. P{\l}oszajczak, Nucl. Phys.
  {\bf A671}, 203 (2000)}.

\bibitem{[Dan98]}
{B.V. Danilin, I.J. Thompson, J.S. Vaagen, and M.V. Zhukov, Nucl. Phys. {\bf
  A632}, 383 (1998)}.

\bibitem{[Nie01]}
{E. Nielsen, D.V. Fedorov, A.S. Jensen, and E. Garrido, Phys. Rep. {\bf 347},
  373 (2001)}.

\bibitem{[Esb99]}
{H. Esbensen and G.F. Bertsch, Phys. Rev. {\bf C} 59, 3240 (1999)}.

\bibitem{[Epp75]}
{D. Eppel and A. Lindner, Nucl. Phys. {\bf A240}, 437 (1975)}.

\bibitem{[Wen87]}
{W.M. Wendler, Nucl. Phys. {\bf A472}, 26 (1987)}.

\bibitem{[Mic02]}
{N. Michel, W. Nazarewicz, M. P{\l}oszajczak, and K. Bennaceur, Phys. Rev.
  Lett. {\bf 89}, 042502 (2002)}.

\bibitem{[Bet02]}
{R. Id Betan, R.J. Liotta, N. Sandulescu, and T. Vertse, Phys. Rev. Lett. {\bf
  89}, 042501 (2002)}.

\bibitem{[Bet02a]}
{R. Id Betan, R.J. Liotta, N. Sandulescu, and T. Vertse,  Phys.
  Rev. C {\bf 67}, 014322 (2003)}.

\bibitem{[Gam28]}
{G. Gamow, Z. Phys. {\bf 51}, 204 (1928); {\bf 52} 510 (1928)}.

\bibitem{[Sie39]}
{A.F.J. Siegert, Phys. Rev. {\bf 56}, 750 (1939)}.

\bibitem{[Ber96a]}
{T. Berggren, Phys. Lett. {\bf B373}, 1 (1996)}.

\bibitem{[Zel60]}
{Y.B. Zel'dovich, JETP (Sov. Phys.) {\bf 39}, 776 (1960)}.

\bibitem{[Hok65]}
{N. Hokkyo, Prog. Theor. Phys. {\bf 33}, 1116 (1965)}.

\bibitem{[Rom68]}
{W.J. Romo, Nucl. Phys. {\bf A116}, 617 (1968)}.

\bibitem{[Ber68]}
{T. Berggren, Nucl. Phys. {\bf A109}, 265 (1968)}.

\bibitem{[Gya71]}
{B. Gyarmati, and T. Vertse, Nucl. Phys. {\bf A160}, 523 (1971)}.

\bibitem{[Gar76]}
{G. Garcia-Calderon and R. Peierls, Nucl. Phys. {\bf A265}, 443 (1976)}.

\bibitem{[Kuk89]}
{V.I.~Kukulin, V.M.~Krasnopol'sky, J.~Hor{\'a}\u{c}ek, {\it Theory of
  Resonances}, Kluwer Academic Publishers (Dordrecht-Boston-London), 1989}.

\bibitem{[Lin93]}
{P. Lind, Phys. Rev. C {\bf 47}, 1903 (1993)}.

\bibitem{[New82]}
{R.G.~Newton, {\it Scattering Theory of Waves and Particles} (Springer-Verlag,
  New York Heidelberg Berlin) 1982}.

\bibitem{[Jos47]}
{R. Jost, Helv. Phys. Acta {\bf 20}, 256 (1947)}.

\bibitem{[Ver87]}
{T. Vertse, P. Curutchet, and R.J. Liotta, Lecture Notes in Physics {\bf 325}
  (Springer Verlag, Berlin 1987), p. 179}.

\bibitem{[Lio96]}
{R.J. Liotta, E. Maglione, N. Sandulescu, and T. Vertse, Phys. Lett. {\bf
  B367}, 1 (1996)}.

\bibitem{[Ver98]}
{T. Vertse, R.J. Liotta, W. Nazarewicz, N. Sandulescu, and A.T. Kruppa, Phys.
  Rev. {\bf C57}, 3089 (1998)}.

\bibitem{[Dus92]}
{G.G. Dussel, R.J. Liotta, H. Sofia, and T.Vertse, Phys. Rev. {\bf C46}, 558
  (1992)}.

\bibitem{[For97]}
{S. Fortunato, A. Insolia, R.J. Liotta, and T. Vertse, Phys. Rev. {\bf C54},
  3279 (1997)}.

\bibitem{[Civ01]}
{O. Civitarese, R.J. Liotta, and T. Vertse, Phys. Rev. {\bf C 64}, 057305
  (2001)}.

\bibitem{[Ver95]}
{T. Vertse, R.J. Liotta, and E. Maglione, Nucl. Phys. {\bf A584}, 13 (1995)}.

\bibitem{[Lin94]}
{P. Lind, R.J. Liotta, E. Maglione, and T. Vertse, Z. Phys. {\bf A347}, 231
  (1994)}.

\bibitem{[Gin84]}
{J.N. Ginocchio, Annals of Physics (NY) {\bf 152}, 203 (1984); {\bf 159}, 467
  (1985)}.

\bibitem{[Gre65]}
{I.M. Green and S.A. Moszkowski, Phys. Rev. {\bf 139}, B790 (1965)}.

\bibitem{[Fir96]}
{R.B. Firestone and V.S. Shirley, {\em Table of Isotopes}, Eighth Edition, Vol.
  I (Wiley, New York)}.

\bibitem{[Gya72]}
{B. Gyarmati, F. Krisztinkovics, and T. Vertse, Phys. Lett. {\bf 41B}, 475
  (1972)}.

\bibitem{[Bol96]}
{C.G. Bollini, O. Civitarese, A.L. De Paoli, and M.C. Rocca, Phys. Lett. {\bf B
  382}, 205 (1996)}.

\bibitem{[Civ99]}
{O. Civitarese, M. Gadella, and R.I. Betan, Nucl. Phys. {\bf A660}, 255
  (1999)}.

\end{thebibliography}

\end{document}